\documentclass[a4paper,12pt]{article}
\usepackage[english]{babel}
\usepackage[utf8]{inputenc}
\usepackage[T1]{fontenc}
\usepackage[margin=2.5cm]{geometry}
\usepackage{amsmath}
\usepackage{amsthm}
\usepackage{amssymb}
\usepackage{graphicx}
\usepackage{float}
\usepackage[hidelinks]{hyperref}
\usepackage{booktabs}
\usepackage{threeparttable}
\usepackage[round]{natbib}
\usepackage{xcolor}
\usepackage{tcolorbox}
\usepackage{upquote}
\usepackage{setspace}
\usepackage[affil-it]{authblk}

\theoremstyle{definition}
\newtheorem{definition}{Definition}
\theoremstyle{plain}
\newtheorem{proposition}{Proposition}

\theoremstyle{remark}

\usepackage{subfig}

\providecommand{\keywords}[1]{\textbf{\textit{Keywords ---}} #1}

\begin{document}

\title{Global and local indicators of spatial connectivity for areal data}

\author[1*]{\'{A}lvaro Briz-Red\'{o}n}
\author[2]{Marco Ruiz-Valderrama}

\affil[1]{Departament d'Estad\'{i}stica i Investigaci\'{o} Operativa, Universitat de Val\`encia, Val\`encia, Spain}
\affil[2]{Facultat de Ci\`encies Matem\`atiques, Universitat de Val\`encia, Val\`encia, Spain}
\affil[*]{Corresponding author (\texttt{alvaro.briz@uv.es})}

\maketitle

\begin{abstract}

Methods for areal data commonly describe spatial structure through global autocorrelation measures or local indicators based on neighboring values. In many applications, however, it is also relevant to know whether high- or low-valued areas form connected spatial structures. Motivated by topological data analysis, we develop global and local indicators of spatial connectivity for observations defined on an areal adjacency graph.

We construct thresholded spatial graphs and use the Betti-0 curve to record their number of connected components. We further derive a local decomposition in which each area's contribution to the Betti-0 curve is expressed as an activation term minus a merging term. Each component fusion contributes to the merging term through a symmetric allocation between the entering area and the neighboring areas through which the fusion occurs. Integrating the merging contributions yields area-level indicators of participation in the connectivity of high- or low-valued regions. Global and local inference is based on random relabeling, with a global test for departures from spatial exchangeability and a conditional test for the local indicators.

We illustrate the framework using COVID-19 incidence in Italian provinces during two epidemic waves. Significant departures toward greater connectivity were found for both high- and low-incidence areas, with the largest discrepancy corresponding to high incidence during the first wave. The proposed indicators complement conventional measures of spatial association and are applicable whenever the spatial coherence of threshold exceedances or deficits is of interest.

\end{abstract}


\newpage

\section{Introduction}

Spatial statistics provides a broad set of tools for studying how observations are organized over a spatial support. In areal data analysis, values are attached to administrative regions, census units, epidemiological surveillance areas, ecological zones, or other polygonal units, while spatial structure is usually represented through a neighborhood graph or spatial-weights matrix. Classical measures such as Moran's $I$ \citep{moran1950} and Geary's $C$ \citep{geary1954} summarize the overall degree of similarity among spatially neighboring observations. Local indicators of spatial association, including LISA \citep{anselin1995}, provide area-specific measures that can be used to identify local clusters and spatial outliers.


These measures are central to applied spatial analysis. In this work, we focus on a complementary aspect of spatial structure: spatial connectivity. Spatial association and spatial connectivity capture different features of a spatial configuration. A map may exhibit strong local similarity while its highest-valued areas are divided into several disconnected foci. Conversely, areas with moderately high values may link more extreme areas into a single connected region. Thus, beyond asking whether neighboring areas tend to have similar values, it is also useful to examine whether high- or low-valued areas form connected structures over the areal graph. This distinction is relevant in many applications. Disease incidence may be concentrated in one connected region or across several separated high-incidence foci; environmental exposure may form coherent patches, corridors, or fragmented hotspots; and ecological degradation, drought, wildfire risk, or socioeconomic disadvantage may occupy connected zones even when local association varies across space. Distinguishing between coherent spatial structures and isolated foci may therefore be important for monitoring, interpretation, and spatial prioritization.

Topological data analysis (TDA) is a field at the intersection of topology, data analysis, and statistics that provides a natural framework for studying connectivity \citep{wasserman2018topological,chazal2021introduction}. In particular, many TDA methods rely on the notion of persistent homology, which describes how topological features evolve across scales \citep{edelsbrunner2002,zomorodian2005,carlsson2009,edelsbrunnerHarer2010}, while stability results and functional summaries such as persistence landscapes support its statistical use \citep{cohenSteiner2007,bubenik2015}. TDA has also been applied to a variety of spatial problems. Examples include goodness-of-fit and related inference for point processes \citep{biscio2019,fendRedenbach2025,eckardtMoradi2026}, methods for random sets and spatial tessellations \citep{gotovacDogasMandaric2025,hirsch2024}, procedures for unsupervised space-time clustering \citep{islambekovGel2019}, and applications to voting data analysis \citep{fengPorter2021}, epidemiology \citep{hickokNeedellPorter2022}, geographical information science \citep{corcoranJones2023}, transcriptomics \citep{limbeckRieck2024}, and spatially structured biological systems \citep{mcdonald2026}.

In the context of areal data, observations are attached to a fixed set of spatial units, with adjacency encoded by a spatial graph. In this setting, thresholding the observations produces induced subgraphs whose organization changes with the threshold. Active areas may remain spatially separated or become connected as less extreme values are included. The zeroth Betti number, or Betti-0, summarizes this evolution by counting the connected components of each thresholded graph. Our framework uses this quantity to describe how high- or low-valued regions connect across thresholds.

In particular, in this paper, we propose global and local indicators of spatial connectivity for areal data. The superlevel analysis examines the spatial organization of high-valued areas by retaining, at each threshold, the areas whose values are at or above that threshold together with the adjacency links between them. Conversely, the sublevel analysis examines the organization of low-valued areas by retaining those at or below the corresponding threshold. As the threshold varies, these constructions generate nested sequences of spatial graphs. The resulting Betti-0 curves record how the number of connected regions changes across thresholds, distinguishing patterns concentrated in a small number of connected regions from configurations made up of several separated foci. Departures from spatial exchangeability are assessed by random relabeling over the fixed graph.

To obtain local information, we derive an activation--merging decomposition of the Betti-0 curve. At each threshold, the net contribution of an area is written as its activation contribution minus its merging contribution. As the threshold is lowered, an area may become active and join one or more components that are already present, with each such fusion reducing the component count by one. We allocate each fusion between the entering area and the neighboring areas through which it occurs, treating equivalent links to the same previously active component symmetrically. Integrating the local contributions across thresholds gives the local indicators of spatial connectivity (LISC), which identify areas that participate strongly in connecting high- or low-valued regions.

Inference is developed at both scales using random relabeling over the fixed graph. The global test assesses whether the observed Betti-0 curve differs more from spatially exchangeable configurations than expected by chance. For local inference, the observed value at the focal area is held fixed while the remaining values are randomly reassigned over the other vertices, so the test asks whether its merging contribution is unusually large given its observed activation value and graph position. We illustrate the method using cumulative COVID-19 incidence in Italian provinces during two epidemic waves and compare the local results with Local Moran's $I$. Although the application is epidemiological, the same framework is intended for environmental and ecological areal data whenever the connectivity of exceedances, deficits, or other threshold-defined patterns is scientifically relevant.

The remainder of the paper is organized as follows. Section~\ref{sec:preliminaries} introduces the graph-based setting and the Betti-0 curve. Section~\ref{sec:methods} presents the global test, the local decomposition, and the conditional local procedure. Section~\ref{sec:application} describes the COVID-19 application, and Section~\ref{sec:discussion} summarizes the main findings and limitations.

\section{Preliminaries}
\label{sec:preliminaries}

We first introduce the graph representation of the areal data and explain how superlevel and sublevel deviations from a reference level are represented through thresholded graphs. We then formalize the nested sequence of graphs obtained as the threshold varies. These constructions provide the setting for the Betti-0 curve and the global and local connectivity measures developed in the next section.

\subsection{Areal data on a spatial graph}
\label{subsec:areal-data}

Let $D$ be a study region partitioned into $n$ non-overlapping areal units. Their adjacency structure is represented by a finite undirected graph $G=(V,E)$, where $V=\{1,\ldots,n\}$ is the set of areal units and $\{i,j\}\in E$ indicates that areas $i$ and $j$ are neighbors according to a chosen rule. Adjacency may be defined by shared boundaries or vertices, distance thresholds, or another scientifically meaningful criterion \citep{brizRedon2022neighborhood}. Throughout the paper, the graph is treated as fixed.

Let $\mathbf{x}=(x_1,\ldots,x_n)$ denote the original scalar-valued observations. We study connectivity relative to a reference level rather than to the numerical origin of the variable, both to give substantive meaning to high- and low-valued regions and to obtain a common formulation of superlevel and sublevel connectivity over positive thresholds. For a reference value $b\in\mathbb{R}$ and a scale $s>0$, define
\[
z_i=\frac{x_i-b}{s},
\qquad i=1,\ldots,n.
\]
The value $b$ is the baseline against which the observations are compared: $z_i>0$ exactly when $x_i>b$, whereas $z_i<0$ when $x_i<b$. It may represent a scientifically meaningful reference or a distributional location such as the mean or median. The scale $s$ controls how distance from this baseline is expressed and therefore sets the units of the thresholds and integrated quantities; choosing $s=1$ retains the original units. Positive entries of $\mathbf{z}$ therefore represent observations above the reference, while positive entries of $-\mathbf{z}$ represent observations below it.

For a vertex subset $U\subseteq V$, the induced subgraph $G[U]$ contains the vertices in $U$ and all edges of $G$ with both endpoints in $U$. A connected component is a maximal set of vertices joined by paths in the graph. These standard graph-theoretic notions are sufficient for the constructions below.

\subsection{Superlevel and sublevel analyses}
\label{subsec:superlevel-sublevel-graphs}

All definitions are first given for the reference-centered vector $\mathbf{z}$. The sublevel analysis is obtained by replacing $\mathbf{z}$ with $-\mathbf{z}$, so the same thresholding and connectivity definitions apply in both directions without introducing a second analysis vector. This convention is used throughout the methodological development and the case study.

\begin{definition}[Superlevel spatial graph]
\label{def:superlevel-sublevel-graphs}
For a threshold $\lambda\in\mathbb{R}$, define
\[
V_\lambda(\mathbf{z})=\{i\in V:z_i\geq\lambda\}
\]
and let
\[
G_\lambda(\mathbf{z})
=
G[V_\lambda(\mathbf{z})]
=
\bigl(V_\lambda(\mathbf{z}),E_\lambda(\mathbf{z})\bigr),
\]
where
\[
E_\lambda(\mathbf{z})
=
\bigl\{\{i,j\}\in E:i,j\in V_\lambda(\mathbf{z})\bigr\}.
\]
When the data vector is clear, we write $V_\lambda$, $G_\lambda$, and $E_\lambda$.
\end{definition}

For $\lambda\geq0$, the superlevel active set can be written in terms of the original observations as
\[
V_\lambda(\mathbf{z})=\{i\in V:x_i\geq b+s\lambda\}.
\]
The sublevel analysis is defined by applying the same superlevel construction to $-\mathbf{z}$,
\[
G_\lambda(-\mathbf{z})
=
G\bigl[\{i\in V:-z_i\geq\lambda\}\bigr]
=
G\bigl[\{i\in V:z_i\leq-\lambda\}\bigr],
\]
which is equivalently the sublevel spatial graph of $\mathbf{z}$ at level $-\lambda$. In terms of the original observations, its active set is
\[
V_\lambda(-\mathbf{z})=\{i\in V:x_i\leq b-s\lambda\}.
\]
Thus, the same non-negative threshold $\lambda$ measures distance from the reference in opposite directions: the superlevel analysis considers areas progressively farther above $b$, and the sublevel analysis considers areas progressively farther below it.

\begin{definition}[Superlevel filtration]
\label{def:superlevel-filtration}
The nested family of graphs
\[
\mathcal F(\mathbf{z})=\{G_\lambda(\mathbf{z}):\lambda\geq0\}
\]
is called the superlevel filtration of $\mathbf{z}$. If $0\leq\lambda_1\leq\lambda_2$, then $V_{\lambda_2}(\mathbf{z})\subseteq V_{\lambda_1}(\mathbf{z})$, so decreasing the threshold can only add areas and their adjacencies. The sublevel filtration is obtained analogously as $\mathcal F(-\mathbf{z})$.
\end{definition}

\begin{figure}[htbp]
    \centering
    \includegraphics[width=\linewidth]{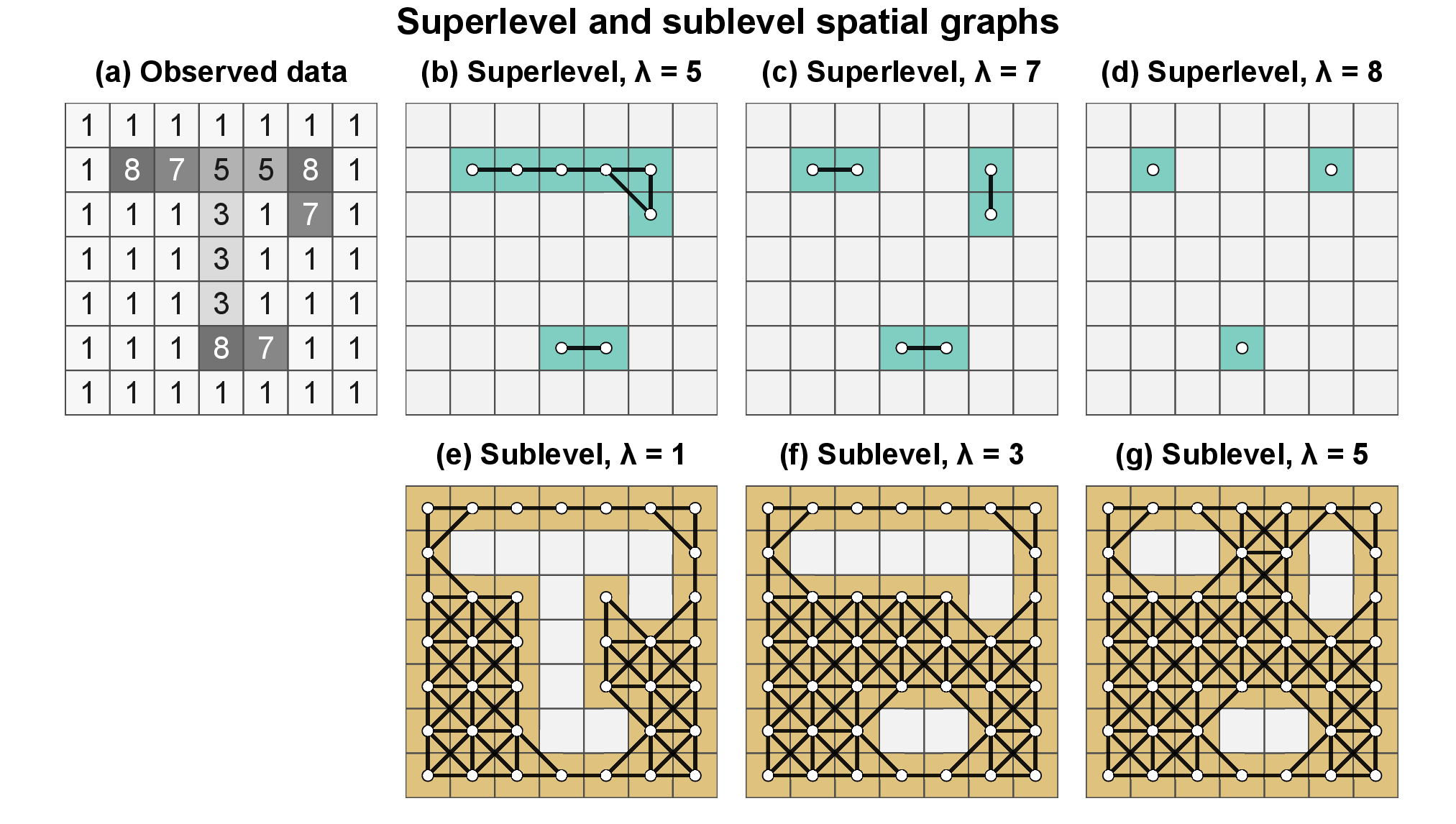}
    \caption{Illustration of superlevel and sublevel thresholded spatial graphs for scalar-valued areal data observed on a regular grid with queen contiguity. Panel (a) shows the observed values. Panels (b)--(d) show superlevel graphs for increasing thresholds, while panels (e)--(g) show the corresponding sublevel construction, equivalently obtained by applying the superlevel rule after reversing the sign of the data. Shaded cells are active areas and black segments are adjacencies between active neighbors.}
    \label{fig:superlevel_sublevel}
\end{figure}

Figure~\ref{fig:superlevel_sublevel} illustrates the construction of both the superlevel and sublevel filtrations for some areal data observed on a regular grid. Several values of $\lambda$ are selected to show how the corresponding thresholded graphs evolve as $\lambda$ varies. For simplicity, the observed data are left untransformed in this example, corresponding to the choice $b=0$ and $s=1$.

\subsection{The Betti-0 curve}
\label{subsec:betti0-preliminaries}

The thresholded graphs describe which areas are active at each level, but a numerical summary is needed to compare their spatial organization across the filtration. We focus on the number of connected components, since it directly distinguishes a single connected region from a configuration formed by several separated regions.

\begin{definition}[Betti-0 curve]
\label{def:betti0-curve}
The Betti-0 curve is
\[
\beta_0(\lambda;\mathbf{z})=c\bigl(G_\lambda(\mathbf{z})\bigr),
\]
where $c(G_\lambda(\mathbf{z}))$ denotes the number of connected components. We use the convention that an empty graph has zero connected components and write $\beta_0(\lambda)$ when the data vector is clear.
\end{definition}

\begin{figure}[htbp]
\centering
\includegraphics[width=\linewidth]{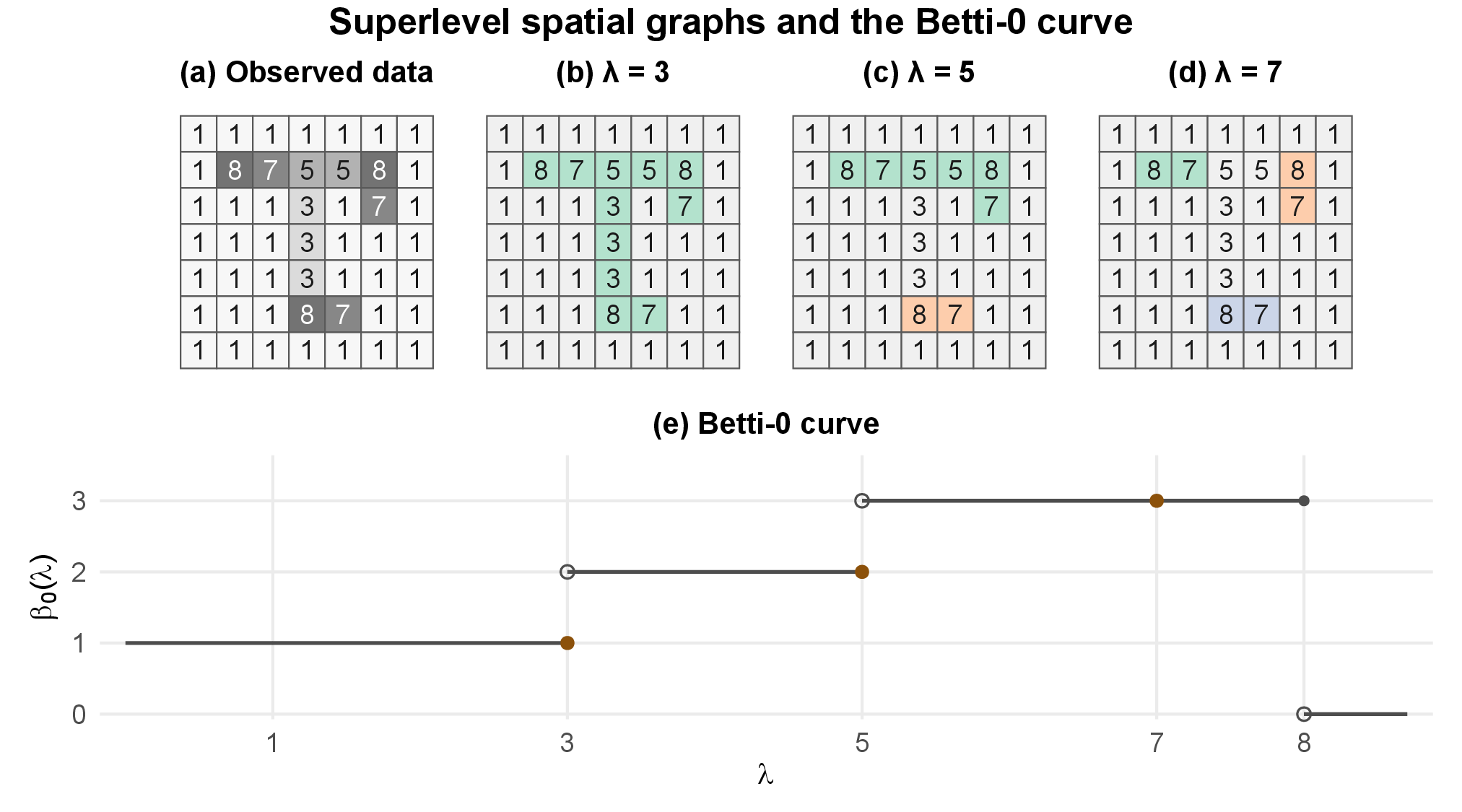}
\caption{Illustration of the Betti-0 curve. Panel (a) shows the observed data. Panels (b)--(d) show the connected components of the superlevel graphs at thresholds $\lambda=3$, $\lambda=5$, and $\lambda=7$, respectively. Panel (e) shows the resulting curve $\beta_0(\lambda)$.}
\label{fig:betti0}
\end{figure}

For a fixed number of active areas, smaller values of $\beta_0$ indicate that those areas form fewer connected regions, while larger values indicate a more fragmented arrangement. As the threshold varies, the curve describes how this connectivity changes across the filtration. Two maps can contain the same number of active areas at a given threshold but have different Betti-0 values if those areas form one connected region in one map and several disconnected regions in the other. Figure~\ref{fig:betti0} illustrates the construction of the Betti-0 curve for the data in Figure~\ref{fig:superlevel_sublevel}.

\section{Methodology}
\label{sec:methods}

All definitions in this section are written for $\mathbf{z}$ and non-negative thresholds. The sublevel analysis follows by replacing $\mathbf{z}$ with $-\mathbf{z}$. We first define a global random-relabeling test based on the complete Betti-0 curve. We then derive the local activation--merging decomposition, integrate it to obtain area-level connectivity indicators, and introduce the corresponding conditional inferential procedure. The main identities are stated as propositions in the corresponding subsections, with their proofs collected in Appendix~\ref{app:proofs}.

\subsection{Global random-relabeling test}
\label{subsec:global-test}

The observed Betti-0 curve does not by itself indicate whether its connectivity pattern is unusual. We therefore use spatial exchangeability over the fixed graph as the null model. The entries of $\mathbf{z}$ are randomly reassigned to the vertices of $G$, preserving their empirical distribution while removing their association with spatial location. The reference $b$ and scale $s$ remain fixed; when they are based on permutation-invariant summaries such as the sample mean and standard deviation, they are unchanged by relabeling.

Let $\mathbf{z}^{(0)}=\mathbf{z}$ denote the observed vector and let $\mathbf{z}^{(1)},\ldots,\mathbf{z}^{(R)}$ be $R$ random relabelings. We use the mean of the $R+1$ curves as a common reference,
\[
\bar\beta_0(\lambda)
=
\frac{1}{R+1}
\sum_{r=0}^R
\beta_0\bigl(\lambda;\mathbf{z}^{(r)}\bigr).
\]
For $k=0,\ldots,R$, we then compute
\[
D^{(k)}(\mathbf{z})
=
\int_0^\infty
\left|
\beta_0\bigl(\lambda;\mathbf{z}^{(k)}\bigr)
-
\bar\beta_0(\lambda)
\right|d\lambda,
\]
and obtain the Monte Carlo $p$-value from the rank of the observed discrepancy among the $R+1$ values,
\[
p(\mathbf{z})
=
\frac{1}{R+1}
\sum_{k=0}^R
I\bigl(D^{(k)}(\mathbf{z})\geq D^{(0)}(\mathbf{z})\bigl),
\]
where $I(\cdot)$ denotes the indicator function.

Small values of $p(\mathbf{z})$ indicate that the observed Betti-0 curve is unusually different from the relabeled configurations. Because $D^{(k)}$ is based on absolute differences, the direction of the departure is determined from the curves themselves: an observed curve below the relabeling mean indicates fewer connected components and greater connectivity, whereas a curve above it indicates greater fragmentation. For graphical presentation, we show the mean and pointwise envelope obtained from the $R$ relabeled curves only.

\subsection{Local decomposition of spatial connectivity}
\label{subsec:lisc-derivation}

The local decomposition follows the order in which areas become active as the superlevel threshold is lowered. We consider the common setting in which the positive entries of $\mathbf{z}$ are distinct, so areas above the reference become active one at a time. This is a natural condition for continuously valued areal variables such as rates, standardized measures, exposures, or model residuals, and it holds in the application below. We restrict the local construction to $\lambda>0$ because the superlevel analysis concerns deviations strictly above the reference. Consequently, only areas with $z_j>0$ enter the superlevel decomposition; areas with $z_j\leq0$ remain inactive over this threshold range and have zero superlevel activation, merging, and net contributions. Applying the same construction to $-\mathbf{z}$ gives the corresponding decomposition for observations below the reference. The single endpoint $\lambda=0$ does not affect the integrated quantities introduced later.

For an area $j$ with $z_j>0$, let
\[
V_j^+=\{h\in V:z_h>z_j\},
\qquad
G_j^+=G[V_j^+].
\]
Immediately before area $j$ becomes active, $G_j^+$ contains the areas that have already entered the filtration. Let $\mathcal C_j$ be the set of connected components of $G_j^+$ containing at least one neighbor of $j$, and write $q_j=|\mathcal C_j|$. Adding $j$ together with its edges to already active neighbors joins it to each of these $q_j$ previously disconnected components. Equivalently, if $j$ is first regarded as a singleton component, the $q_j+1$ components involved are replaced by one. Hence, the activation of $j$ produces exactly $q_j$ non-redundant component fusions and changes the number of connected components by $1-q_j$. Thus, $q_j=0$ means that $j$ creates a new connected component, $q_j=1$ means that it attaches to one existing component without changing the component count, and $q_j\geq2$ means that its entry merges previously disconnected components.

Each fusion described above contributes one unit to the reduction in the number of connected components. To obtain a local decomposition, we distribute this unit among the areas involved in the fusion. We assign one half to the entering area and the remaining half to the already active side of the connection. If several neighbors of the entering area belong to the same active component, they provide alternative links to that component, so this second half is divided equally among them.

More precisely, let $j$ denote the area whose entry creates the fusion and let $i$ denote an area receiving part of its contribution. For $C\in\mathcal C_j$, define
\[
N_j(C)=\{h\in C:\{j,h\}\in E\},
\qquad
k_j(C)=|N_j(C)|.
\]
The contribution assigned to area $i$ from the fusion between $j$ and $C$ is then defined as
\[
\omega_{j\to i}(C)
=
\begin{cases}
\frac{1}{2}, & i=j,\\[3pt]
\frac{1}{2k_j(C)}, & i\in N_j(C),\\[3pt]
0, & \text{otherwise}.
\end{cases}
\]
For every fusion $(j,C)$, these weights sum to one:
\[
\sum_{i\in V}\omega_{j\to i}(C)=1.
\]
This normalization is component-specific: the weights sum to one separately for each $C\in\mathcal C_j$. Consequently, if area $j$ touches $q_j$ previously active components, its entry contributes $q_j$ merging units in total, with one half of each unit assigned to $j$ and the remaining half distributed among its neighbors in the corresponding component.

\begin{definition}[Local activation, merging and net contribution]
\label{def:local-terms}
For area $i$ and threshold $\lambda>0$, define
\[
a_i(\lambda;\mathbf{z})=I(z_i\geq\lambda),
\]
\[
m_i(\lambda;\mathbf{z})
=
\sum_{\substack{j:z_j\geq\lambda}}
\sum_{C\in\mathcal C_j}
\omega_{j\to i}(C),
\]
and
\[
\ell_i(\lambda;\mathbf{z})
=
a_i(\lambda;\mathbf{z})-m_i(\lambda;\mathbf{z}).
\]
When the data vector is clear, we omit $\mathbf{z}$ from the notation.
\end{definition}

The activation term $a_i(\lambda)$ records whether area $i$ is active, while $m_i(\lambda)$ accumulates the shares assigned to area $i$ from all component fusions that have occurred at or above threshold $\lambda$. When $i$ enters, it receives one half-unit from each of the $q_i$ components that it joins, giving an initial contribution $q_i/2$. It may subsequently receive additional merging shares when a neighboring area with a smaller value enters and connects to the already active component containing $i$. Thus, $q_i$ characterizes only the fusions generated by the entry of $i$, whereas $m_i(\lambda)$ summarizes its overall participation in fusions throughout the filtration. Accordingly, larger values of $m_i(\lambda;\mathbf{z})$ indicate greater participation by area $i$ in the fusions that determine the connectivity of the superlevel set at threshold $\lambda$; $m_i(\lambda;-\mathbf{z})$ has the analogous interpretation for the sublevel set. Finally, the net term \(\ell_i(\lambda)\) represents the balance between activation and merging at that threshold and may become negative when merging exceeds activation.

The following proposition shows that these local activation and merging contributions reproduce the Betti-0 curve exactly.

\begin{proposition}[Activation--merging representation of the Betti-0 curve]
\label{prop:activation-merging-decomposition}
For every $\lambda>0$,
\[
|V_\lambda(\mathbf{z})|
=
\sum_{i\in V}a_i(\lambda;\mathbf{z}),
\qquad
\sum_{\substack{j:z_j\geq\lambda}}q_j
=
\sum_{i\in V}m_i(\lambda;\mathbf{z}),
\]
and
\[
\beta_0(\lambda;\mathbf{z})
=
|V_\lambda(\mathbf{z})|
-
\sum_{\substack{j:z_j\geq\lambda}}q_j
=
\sum_{i\in V}a_i(\lambda;\mathbf{z})
-
\sum_{i\in V}m_i(\lambda;\mathbf{z})
=
\sum_{i\in V}\ell_i(\lambda;\mathbf{z}).
\]
\end{proposition}

Beyond providing a local representation, the same decomposition links the global and local views of connectivity. For $k=0,\ldots,R$, let $m_i(\lambda;\mathbf{z}^{(k)})$ denote the local merging contributions for configuration $k$. Since every relabeling preserves the multiset of values, the total number of active areas is identical across all $R+1$ configurations at any fixed threshold. Consequently, differences between their Betti-0 curves can only arise from differences in total merging. This observation gives the following decomposition of the global curve discrepancy.

\begin{proposition}[Random-relabeling decomposition]
\label{prop:random-relabel}
For every $k=0,\ldots,R$ and threshold $\lambda>0$,
\[
\beta_0\bigl(\lambda;\mathbf{z}^{(k)}\bigr)
-
\bar\beta_0(\lambda)
=
-
\left[
\sum_{i\in V}m_i\bigl(\lambda;\mathbf{z}^{(k)}\bigr)
-
\frac{1}{R+1}
\sum_{r=0}^R
\sum_{i\in V}m_i\bigl(\lambda;\mathbf{z}^{(r)}\bigr)
\right].
\]
\end{proposition}

Hence, differences between a Betti-0 curve and the common random-relabeling reference are entirely due to differences in total merging. The same result applies to the sublevel analysis after replacing $\mathbf{z}$ with $-\mathbf{z}$.

The local contributions also satisfy simple bounds at each threshold. For every area $i$ and $\lambda>0$,
\[
0\leq m_i(\lambda)\leq\frac{d_i^G}{2},
\]
where $d_i^G$ is the degree of area $i$ in the original adjacency graph. If area $i$ is inactive, then $m_i(\lambda)=\ell_i(\lambda)=0$; if it is active, $\ell_i(\lambda)=1-m_i(\lambda)$. The degree bound follows because each neighbor can account for at most one half-unit of the merging allocation received by an area. The net contribution $\ell_i(\lambda)$ can be negative and should not be interpreted as a local Betti number, whereas $m_i(\lambda)$ measures participation in fusions that reduce the number of connected components.

\subsection{Integrated local indicators of spatial connectivity}
\label{subsec:lisc}

The threshold-specific quantities above describe the role of each area at a particular threshold. To obtain one local summary for mapping, comparison, and inference, we integrate these contributions over all positive thresholds,
\[
A_i(\mathbf{z})
=
\int_0^\infty a_i(\lambda;\mathbf{z})\,d\lambda,
\qquad
M_i(\mathbf{z})
=
\int_0^\infty m_i(\lambda;\mathbf{z})\,d\lambda,
\qquad
L_i(\mathbf{z})
=
A_i(\mathbf{z})-M_i(\mathbf{z}).
\]
The activation term has the closed form $A_i(\mathbf{z})=(z_i)_+$, where $(a)_+=\max(a,0)$. The integrated merging term $M_i(\mathbf{z})$ accumulates all merging contributions assigned to area $i$ across the threshold range over which they are present. It therefore reflects both fusions generated when $i$ becomes active and contributions received from later-entering neighboring areas.

The integrated local quantities retain an exact relation with the Betti-0 curve. Integrating Proposition~\ref{prop:activation-merging-decomposition} over positive thresholds gives a direct connection between the global curve and the local contributions. Define $B(\mathbf{z}) = \int_0^\infty \beta_0(\lambda;\mathbf{z})\,d\lambda$. Because \(\beta_0(\lambda;\mathbf{z})\) is piecewise constant and vanishes above \(\max_i(z_i)_+\), this integral is finite. It then follows that

$$ B(\mathbf{z}) = \sum_{i\in V}A_i(\mathbf{z}) - \sum_{i\in V}M_i(\mathbf{z}) = \sum_{i\in V}L_i(\mathbf{z}). $$

The quantity \(B(\mathbf{z})\) is introduced here to express this exact integrated local-to-global relation. The global random-relabeling test itself is based instead on the discrepancy \(D(\mathbf{z})\) defined above.

We refer to $M_i(\mathbf{z})$ as the local indicator of spatial connectivity (LISC). Large values of $M_i(\mathbf{z})$ indicate strong cumulative participation in the fusions through which high-valued superlevel regions become connected across thresholds, whereas large values of $M_i(-\mathbf{z})$ indicate the analogous role for low-valued sublevel regions. These magnitudes describe participation in spatial connectivity; the conditional procedure below assesses whether that participation is unusually large given the area's observed activation value and graph position. The net term $L_i(\mathbf{z})$ is the algebraic local contribution to integrated Betti-0 and becomes negative when merging exceeds activation.

\subsection{Local conditional random-relabeling test}
\label{subsec:local-test}

The magnitude of $M_i(\mathbf{z})$ depends partly on $z_i$, because an area with a larger positive value remains active over a wider threshold range and has more opportunity to participate in component fusions. We therefore use a conditional random-relabeling test that keeps the focal value and graph location fixed. For area $i$, let $\mathbf{z}^{(0,i)}=\mathbf{z}$. For $r=1,\ldots,R$, the vector $\mathbf{z}^{(r,i)}$ keeps $z_i$ at vertex $i$ and randomly permutes the values $\{z_j:j\neq i\}$ over the remaining vertices.

For every conditional relabeling, the complete local decomposition is recomputed. A lower-tail test for the net contribution is
\[
p_i(\mathbf{z})
=
\frac{1}{R+1}
\sum_{r=0}^{R}
I\biggl(L_i\bigl(\mathbf{z}^{(r,i)}\bigr)
\leq
L_i(\mathbf{z})\biggl).
\]
The observed configuration is therefore explicitly included among the $R+1$ values. Because $z_i$ remains fixed, $A_i(\mathbf{z}^{(r,i)})=A_i(\mathbf{z})$ for every $r$, and the same $p$-value can be written as the upper-tail test
\[
p_i(\mathbf{z})
=
\frac{1}{R+1}
\sum_{r=0}^{R}
I\biggl(
M_i\bigl(\mathbf{z}^{(r,i)}\bigr)
\geq
M_i(\mathbf{z})
\biggl).
\]
Small values therefore identify areas whose merging contribution is unusually large given their observed activation value and fixed position in the graph. The sublevel analysis applies the same procedure to $-\mathbf{z}$. Since one test is carried out for each area, we report both unadjusted and Benjamini--Hochberg adjusted $p$-values \citep{benjamini1995}.

\subsection{Classification of local spatial connectivity profiles}
\label{subsec:classification}

The continuous indicators and conditional $p$-values contain the main local information, but maps of individual quantities do not always provide a compact description of the different local roles. For visualization, we therefore combine activation, merging, and conditional significance into a small set of activation--connectivity profiles. Let $\eta\in(0,1/2)$ denote an empirical tail proportion and define
\[
c_\eta(\mathbf{z})
=
\max\bigl\{0,Q_{1-\eta}(\mathbf{z})\bigr\},
\]
where $Q_{1-\eta}(\mathbf{z})$ is the empirical $(1-\eta)$-quantile. The truncation at zero ensures that only areas on the active side of the reference can be considered highly activated.

For the superlevel analysis and a significance level $\alpha$, area $i$ is classified as a \textit{core connector} when
\[
z_i>0,
\qquad
z_i\geq c_\eta(\mathbf{z}),
\qquad
p_i(\mathbf{z})<\alpha,
\qquad
L_i(\mathbf{z})\leq0;
\]
as a \textit{bridge connector} when
\[
0<z_i<c_\eta(\mathbf{z})
\qquad\text{and}\qquad
p_i(\mathbf{z})<\alpha;
\]
and as a \textit{peak connector} when
\[
z_i>0,
\qquad
z_i\geq c_\eta(\mathbf{z}),
\qquad
p_i(\mathbf{z})<\alpha,
\qquad
L_i(\mathbf{z})>0.
\]
Areas that do not satisfy these conditions are classified as not significant. Core and peak connectors both have high activation values and show significant merging, but they differ in whether merging is large enough to offset activation in the integrated decomposition. Bridge connectors have more moderate activation values together with significant conditional merging.

For the sublevel analysis, the same rules are applied after replacing $\mathbf{z}$ with $-\mathbf{z}$ and $c_\eta(\mathbf{z})$ with $c_\eta(-\mathbf{z})$. The peak-connector label is then replaced by \textit{trough connector}, since high activation in $-\mathbf{z}$ corresponds to observations far below the reference level. These profiles are descriptive labels that incorporate inferential evidence and should not be interpreted as causal classes. In the application we use $\eta=0.15$ and define profiles from unadjusted $p$-values, while reporting multiplicity-adjusted results separately.

\subsection{Software and implementation}
\label{subsec:software}

The proposed global and local connectivity procedures are implemented in the R package \texttt{spatconnect}, available at \url{https://github.com/albrizre/spatconnect}. Spatial geometries were handled with the \texttt{sf} package \citep{pebesma2018}, while neighborhood structures and Local Moran's $I$ were obtained with \texttt{spdep} \citep{bivandWong2018}. Figures were produced mainly with \texttt{ggplot2} \citep{wickham2016}. The \texttt{spatconnect} repository also contains the code required to reproduce the Italian COVID-19 case study.

\section{Case study: COVID-19 incidence in Italian provinces}
\label{sec:application}

We illustrate the methodology using cumulative COVID-19 incidence in Italian provinces during two epidemic waves. The aim is to compare the spatial organization of high- and low-incidence areas across the two periods and to show how the local connectivity indicators differ from conventional local spatial association. Local Moran's $I$ is used as a benchmark throughout the application.

\subsection{Data, spatial support and study periods}
\label{subsec:data-spatial-support}

The data come from the COVID-19 European Regional Tracker \citep{naqvi2021}, which provides harmonized regional time series for European countries. We use the Italian NUTS-3 data and the corresponding population denominators, together with the NUTS geometries from Eurostat/GISCO \citep{eurostatGISCO}, to calculate cumulative incidence per 100,000 inhabitants for each province-level region. For simplicity, these NUTS-3 regions are referred to as provinces.

The spatial support contains 106 provinces. Queen contiguity is used to define adjacency, so two provinces are neighbors if their polygons share a boundary segment or a vertex. The resulting undirected graph contains 237 edges and has an average vertex degree of 4.47. The full adjacency graph comprises three separate geographic blocks because the spatial support includes insular regions that are not contiguous with mainland Italy. The first wave covers 24 February to 3 May 2020, corresponding to the early national outbreak, while the second wave covers 1 October to 31 December 2020, corresponding to the autumn--winter resurgence.

For each wave, let $\mathbf{x}$ be the vector of cumulative incidence rates. We use the within-wave mean as the reference, $b=\bar x$, and the within-wave standard deviation as the scale, $s=s_x$, so
\[
z_i=\frac{x_i-\bar x}{s_x}.
\]
The superlevel analysis uses $\mathbf{z}$ and the sublevel analysis uses $-\mathbf{z}$. Thus, zero is the within-wave mean and one threshold unit corresponds to one within-wave standard deviation. We use 1999 random relabelings for both the global connectivity test and each conditional local test. Local Moran's $I$ is calculated on the same queen-contiguity graph with row-standardized weights and 1999 conditional permutations, using two-sided unadjusted $p$-values.

\subsection{Descriptive analysis}
\label{subsec:descriptive-incidence}

\begin{figure}[htbp]
    \centering
    \subfloat[]{\includegraphics[width=6.5cm,angle=0]{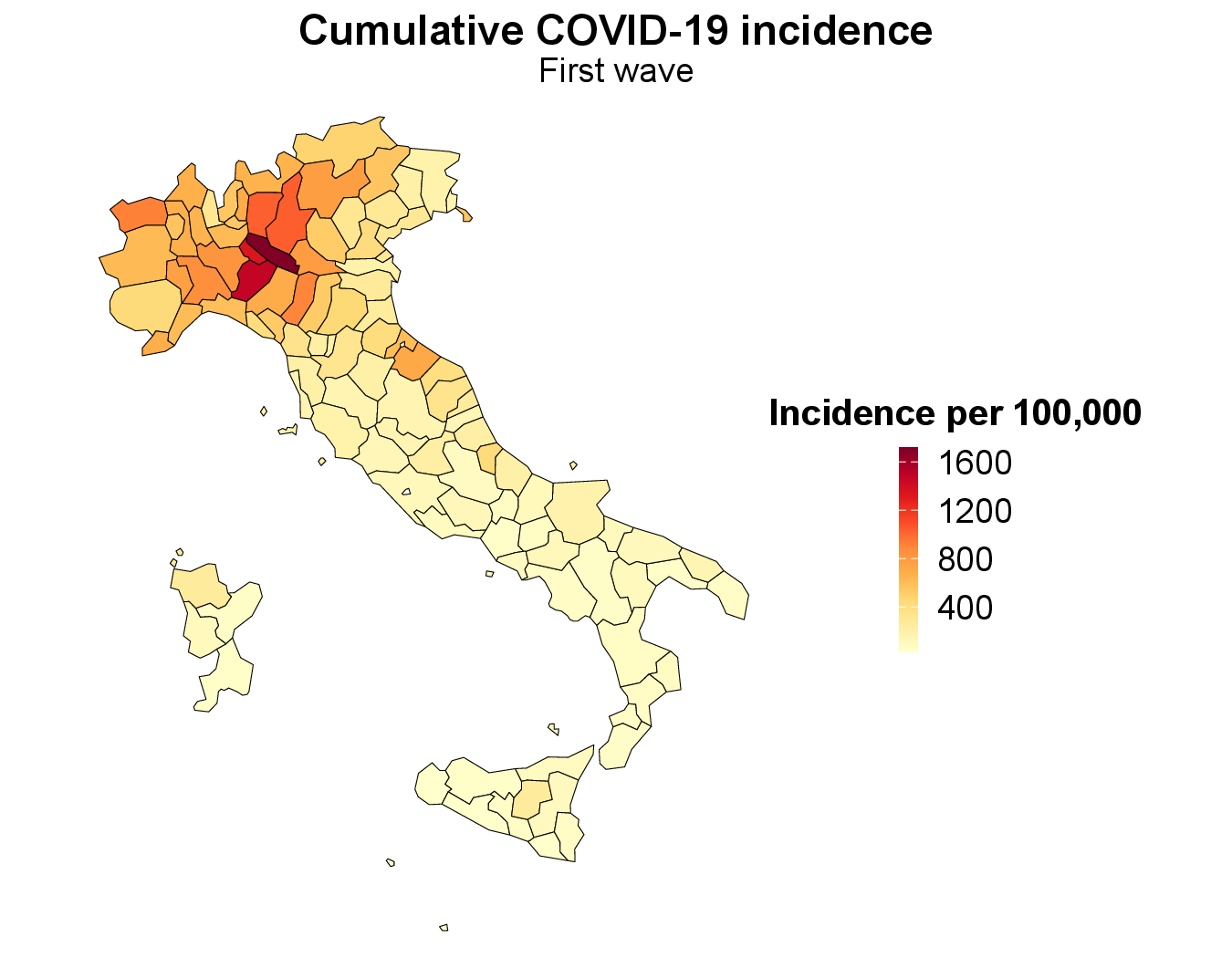}}
    \subfloat[]{\includegraphics[width=6.5cm,angle=0]{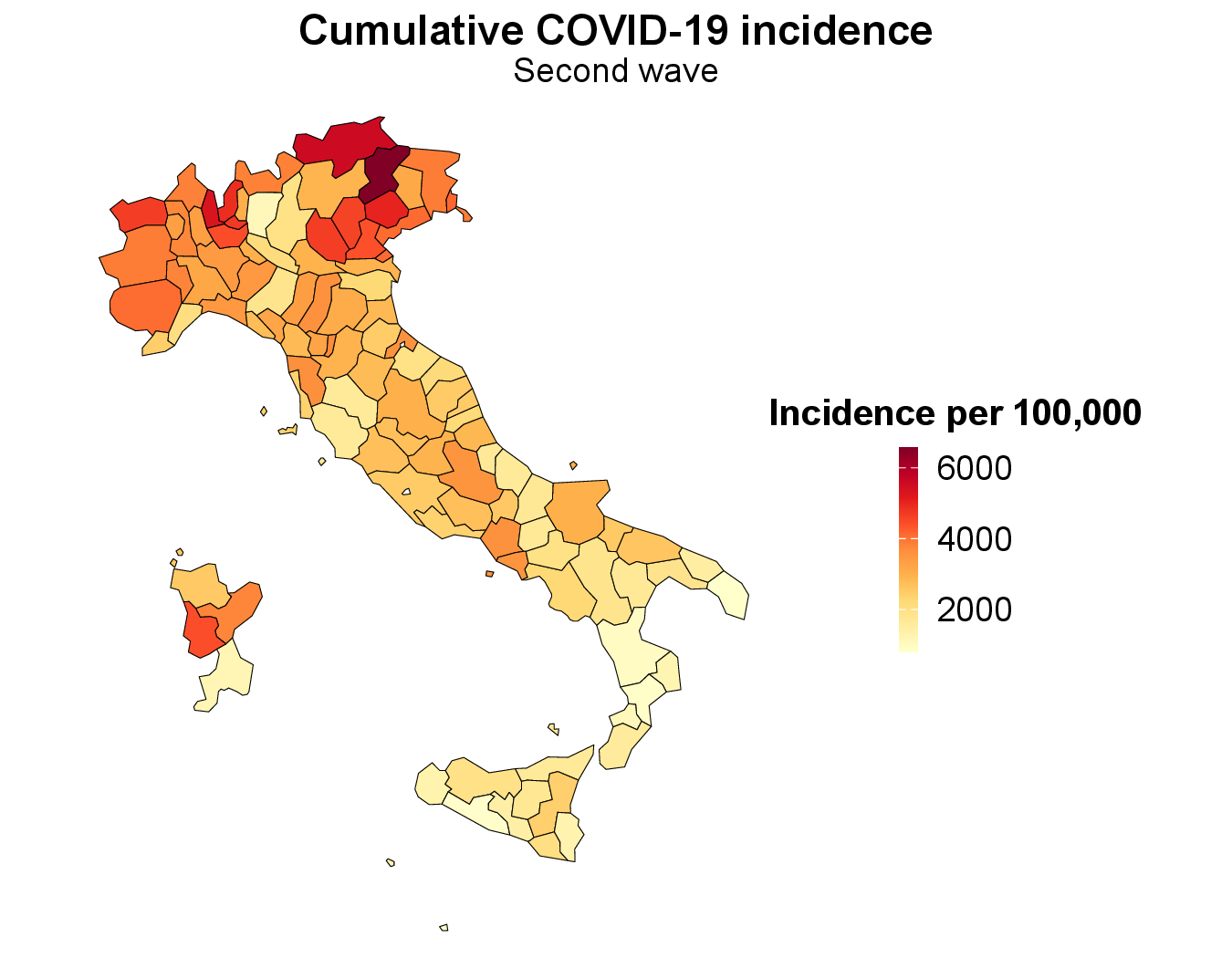}}
    \caption{Cumulative COVID-19 incidence per 100,000 inhabitants in Italian provinces during the two epidemic waves. Panel (a) shows the first wave, from 24 February 2020 to 3 May 2020, and panel (b) shows the second wave, from 1 October 2020 to 31 December 2020.}
    \label{fig:descriptive}
\end{figure}

Before applying the connectivity measures, we examine the spatial distribution of incidence in each wave. Figure~\ref{fig:descriptive} shows cumulative incidence during the two periods. The first wave has a strongly uneven spatial pattern, with the highest values concentrated in Northern Italy, especially in the north-west and the Po Valley. Most central and southern provinces, including the islands, show much lower incidence. The second wave is less geographically concentrated: high values remain present in the north but extend to additional central, southern, and insular provinces. Since the study periods differ in duration and occurred under different surveillance conditions, we do not compare their raw incidence levels directly. The subsequent analysis instead focuses on the within-wave spatial organization of the standardized values.

\begin{figure}[htbp]
    \centering
    \includegraphics[width=15cm,angle=0]{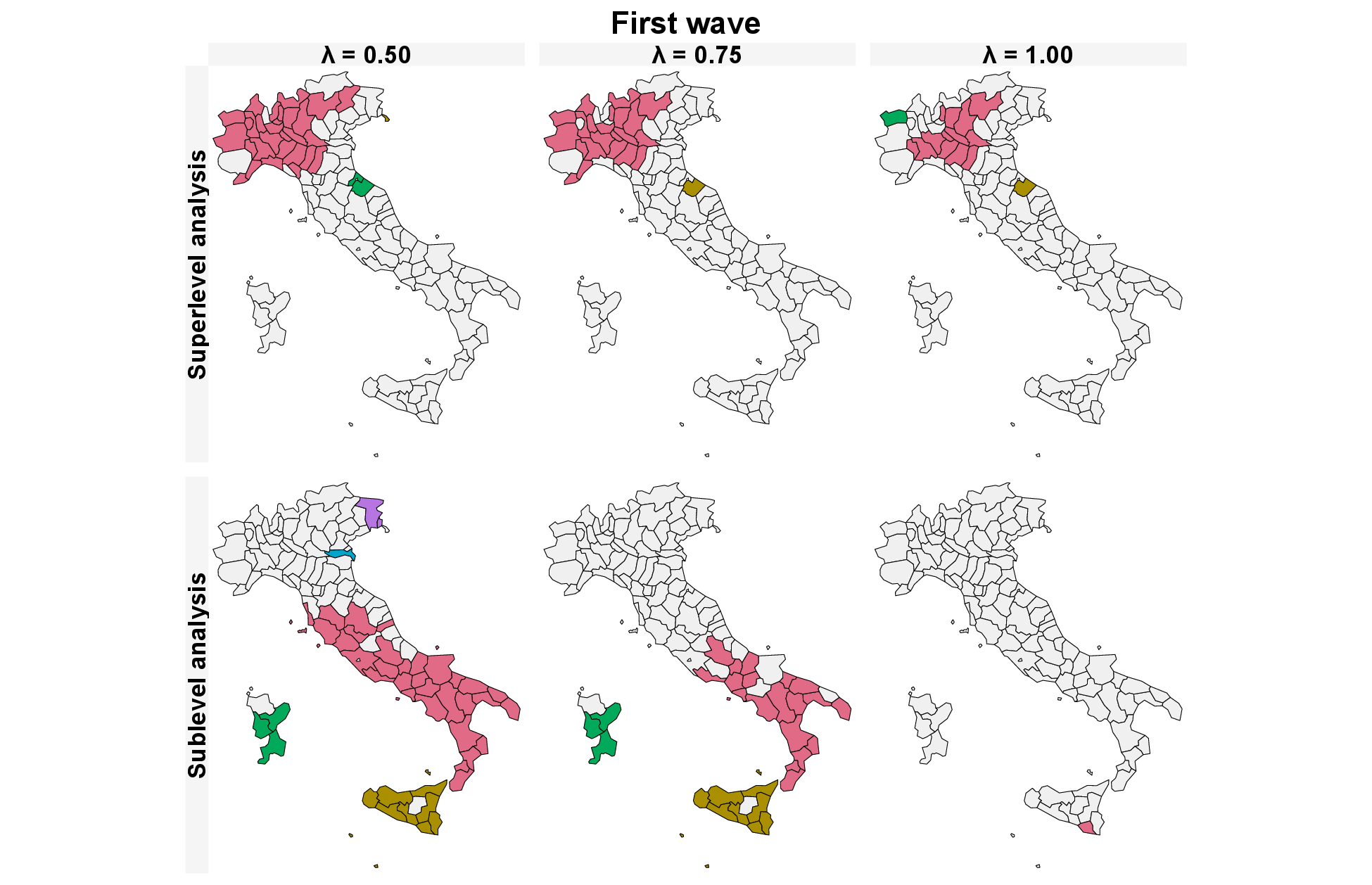}
    \caption{Connected components of the thresholded incidence graph during the first epidemic wave. Columns correspond to three threshold values $\lambda$, and rows correspond to the superlevel analysis based on $\mathbf{z}$ and the sublevel analysis based on $-\mathbf{z}$. Within each panel, provinces belonging to the same connected component share the same color, while inactive provinces are shown in gray.}
    \label{fig:first-wave-components}
\end{figure}

Figure~\ref{fig:first-wave-components} complements these maps by showing the connected components obtained during the first wave at three representative threshold values. The top row corresponds to the superlevel analysis based on $\mathbf{z}$, and the bottom row to the sublevel analysis based on $-\mathbf{z}$. At higher thresholds, the superlevel active set is concentrated in a small number of northern components; as $\lambda$ is lowered, additional provinces become active and these components merge into a broader northern structure. The sublevel analysis shows a different organization, with low-incidence areas forming connected structures mainly across central and southern Italy and the islands. The panels provide a direct spatial illustration of the connectivity changes summarized across all thresholds by the Betti-0 curve.

\subsection{Global spatial connectivity}
\label{subsec:application-global}

\begin{figure}[htbp]
    \centering
    \subfloat[]{\includegraphics[width=8cm,angle=0]{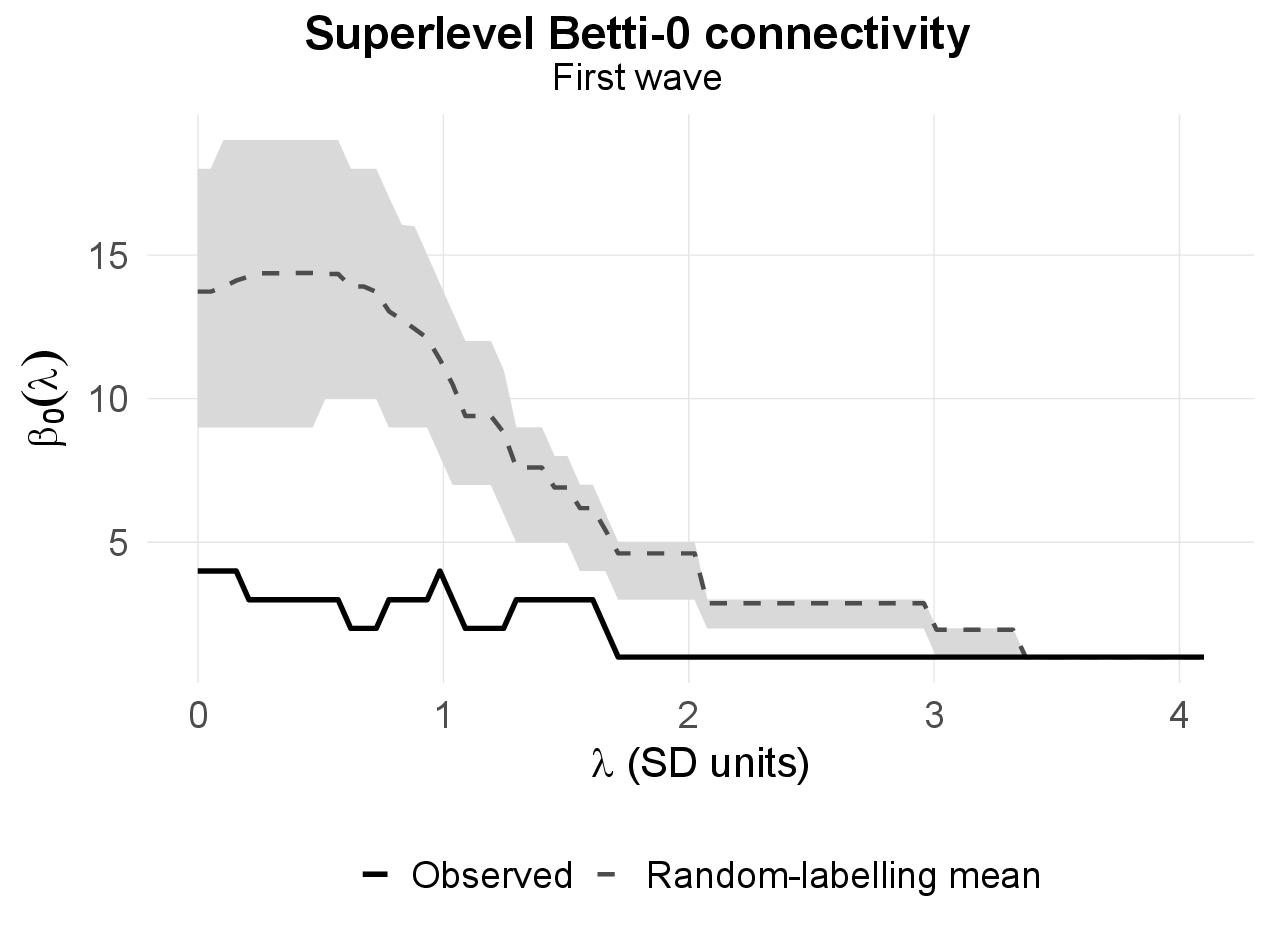}}
    \subfloat[]{\includegraphics[width=8cm,angle=0]{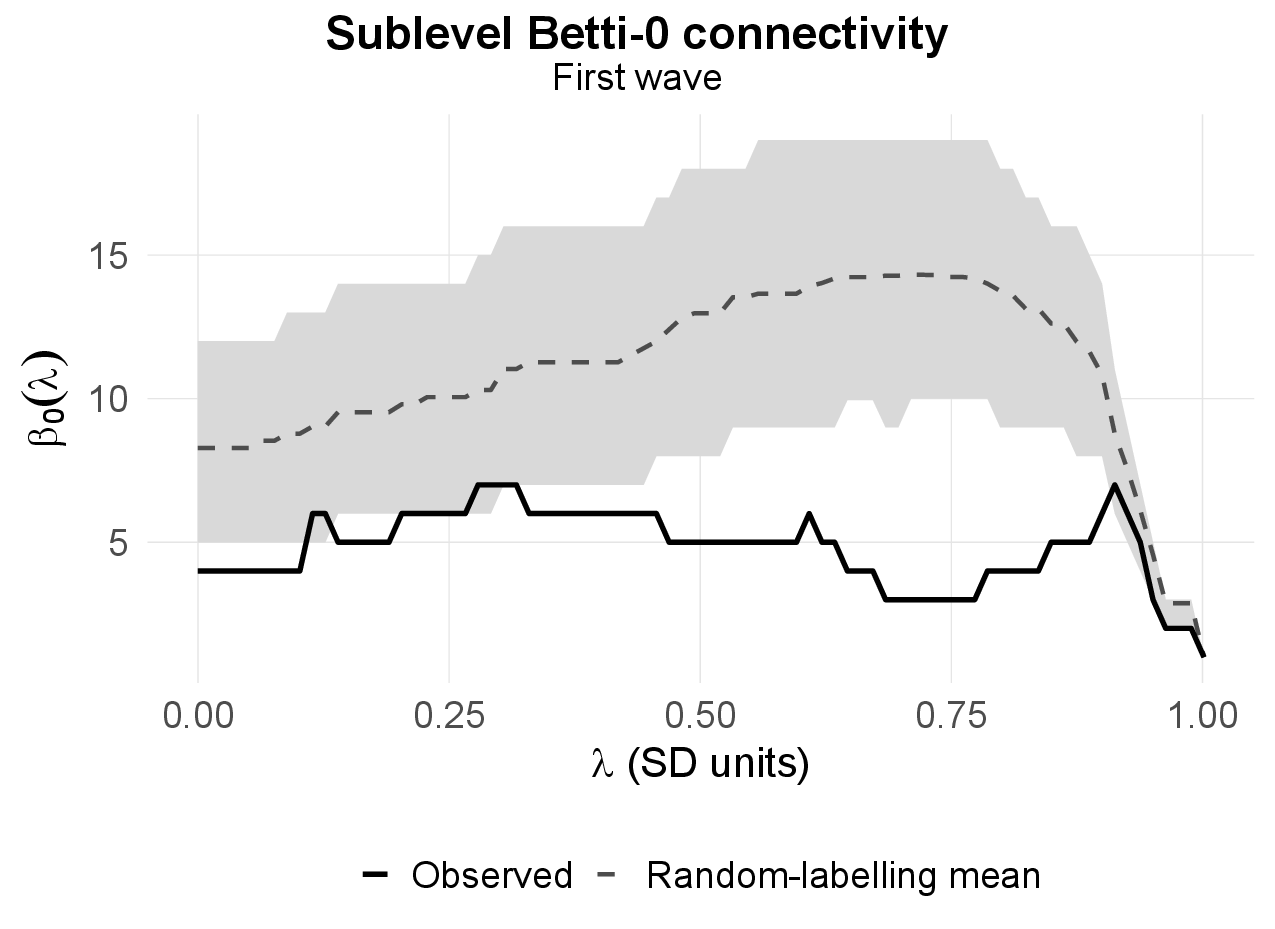}}\\
    \subfloat[]{\includegraphics[width=8cm,angle=0]{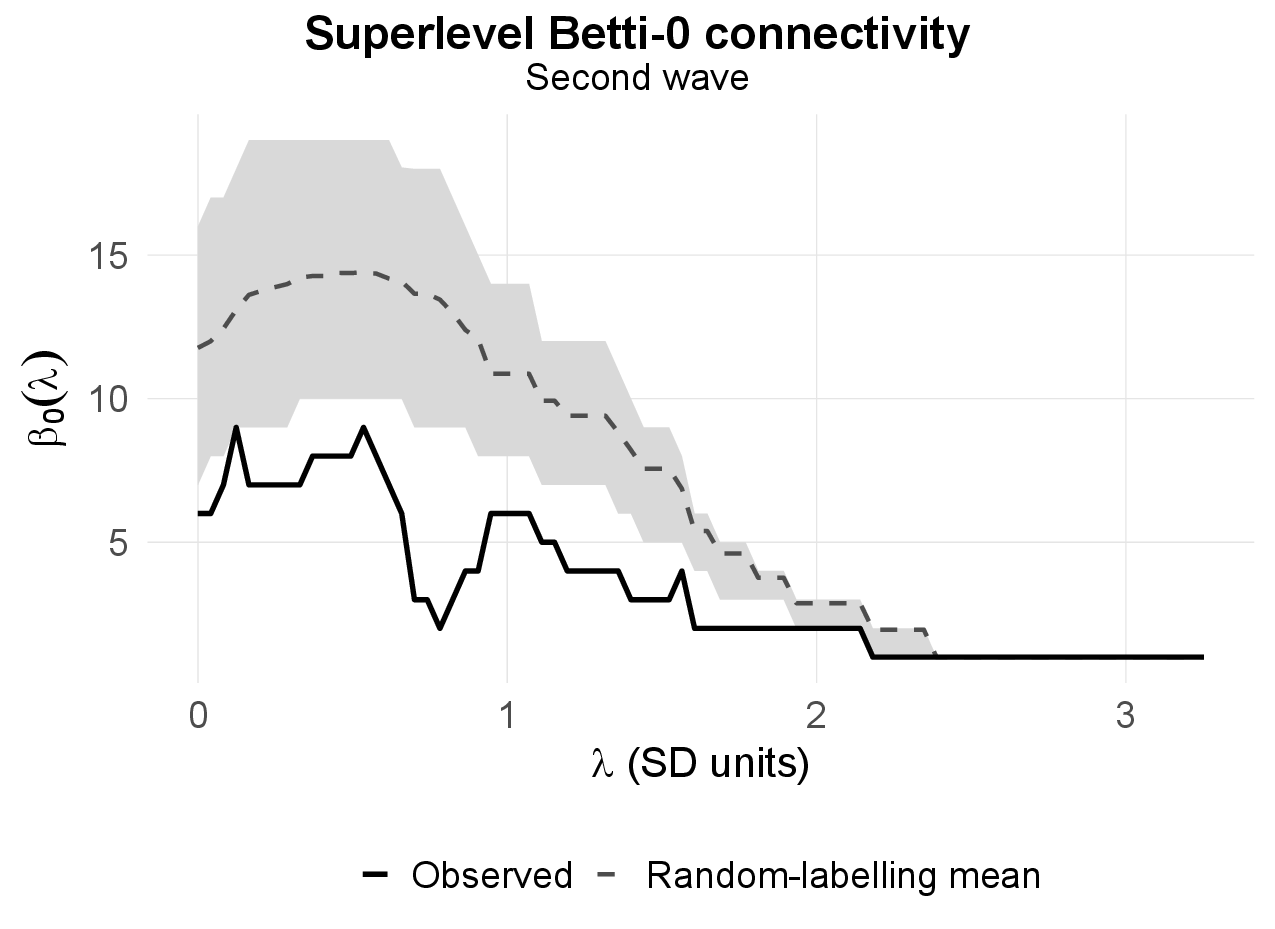}}
    \subfloat[]{\includegraphics[width=8cm,angle=0]{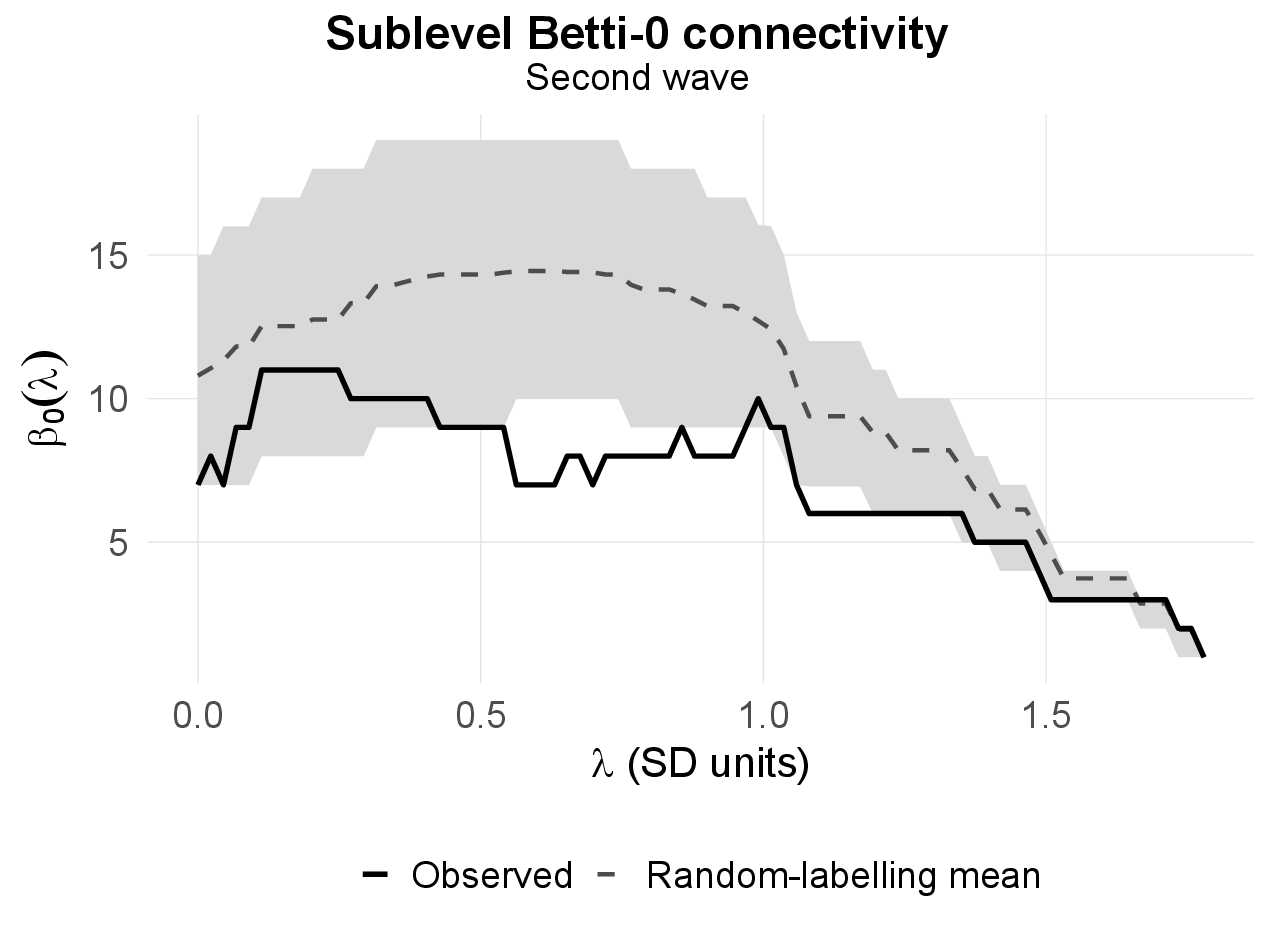}}
    \caption{Betti-0 curves for the two epidemic waves. Panels (a) and (c) show the superlevel analyses based on $\mathbf{z}$ for the first and second waves, respectively. Panels (b) and (d) show the corresponding sublevel analyses based on $-\mathbf{z}$. Solid lines are the observed curves, dashed lines are the random-relabeling mean curves, and shaded regions are pointwise 95\% random-relabeling envelopes.}
    \label{fig:global-betti0-waves}
\end{figure}

We first examine whether the overall connectivity of high- and low-incidence provinces differs from what would be expected under spatial exchangeability. Figure~\ref{fig:global-betti0-waves} compares the observed Betti-0 curves with their random-relabeling references. In the superlevel analysis, the observed curves lie below the relabeling means over broad threshold ranges in both waves. High-incidence provinces therefore form fewer connected components than expected under spatial exchangeability. The global test gives $D^{(0)}(\mathbf{z})=17.65$ with $p(\mathbf{z})=0.0005$ in the first wave and $D^{(0)}(\mathbf{z})=11.19$ with $p(\mathbf{z})=0.0005$ in the second. Since the discrepancy statistic is based on absolute differences, its direction is determined from the curves themselves; in both cases the significant departure corresponds to greater connectivity.

The superlevel discrepancy is larger in the first wave, which agrees with the strong concentration of high incidence in Northern Italy. The sublevel analysis also shows non-random connectivity. For the first wave, $D^{(0)}(-\mathbf{z})=6.24$ with $p(-\mathbf{z})=0.0005$, while for the second wave $D^{(0)}(-\mathbf{z})=5.93$ with $p(-\mathbf{z})=0.009$. The observed sublevel curves again lie mainly below the random-relabeling means. Thus, comparatively low-incidence provinces also form more connected spatial structures than expected under the null model, although these departures are weaker than the first-wave superlevel signal.

Overall, significant departures from spatial randomness were found for both superlevel and sublevel connectivity. The strongest global departure occurs for high incidence during the first wave, while the second wave retains significant but less pronounced superlevel connectivity. These differences are interpreted descriptively rather than as a formal comparison between epidemic waves.

\subsection{Local spatial connectivity}
\label{subsec:local-connectivity-results}

\begin{figure}[htbp]
    \centering
    \subfloat[]{\includegraphics[width=8cm,angle=0]{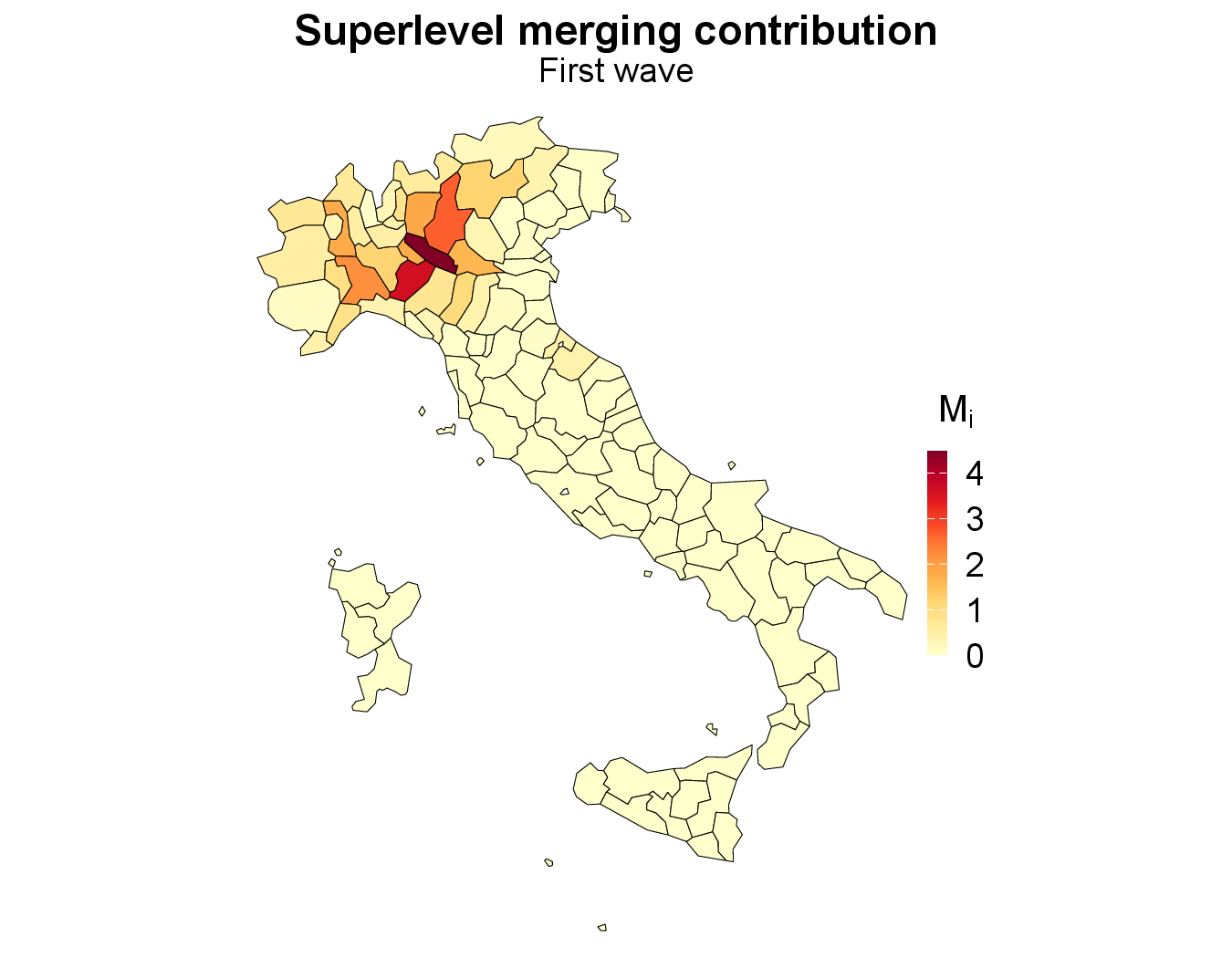}}
    \subfloat[]{\includegraphics[width=8cm,angle=0]{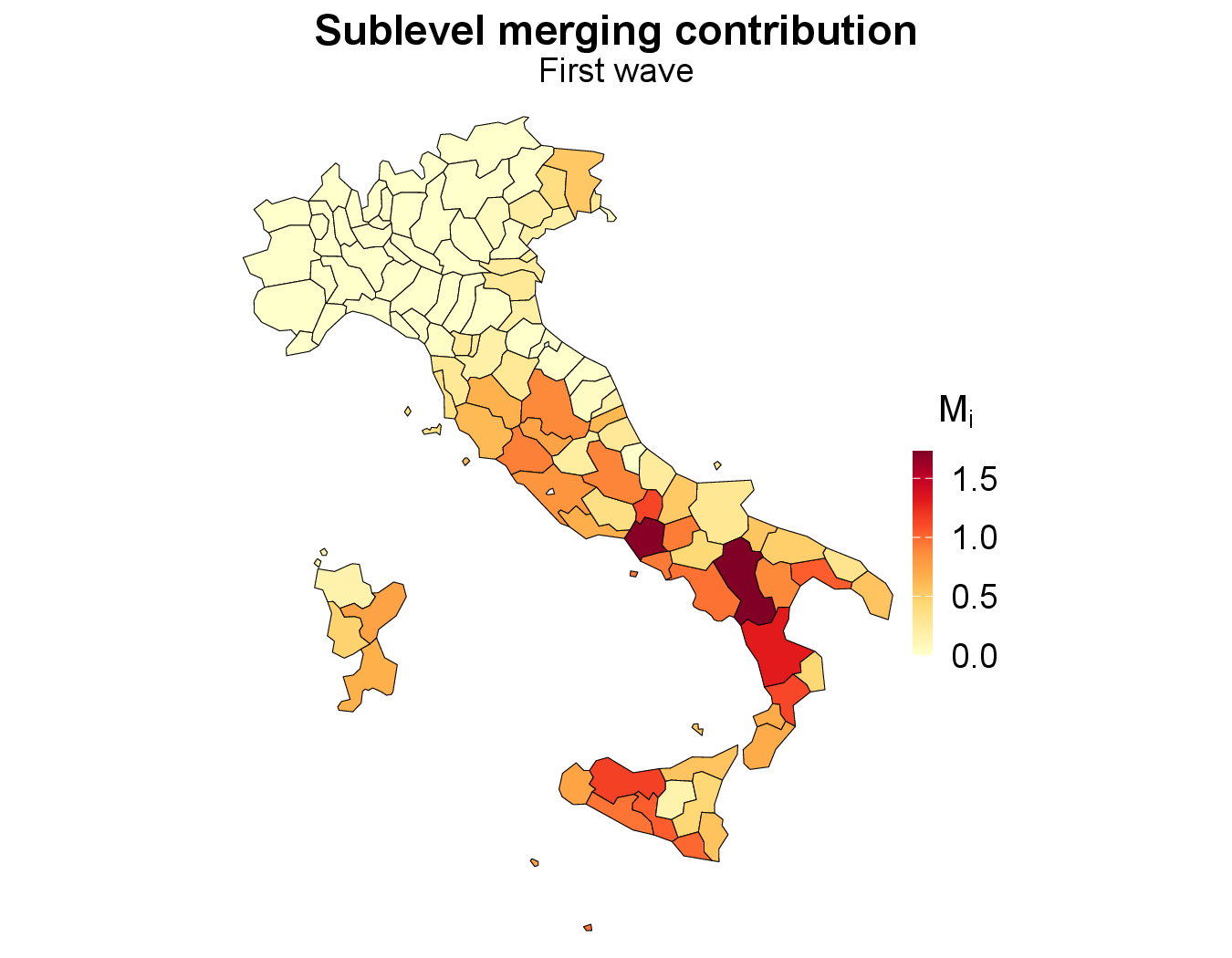}}\\
    \subfloat[]{\includegraphics[width=8cm,angle=0]{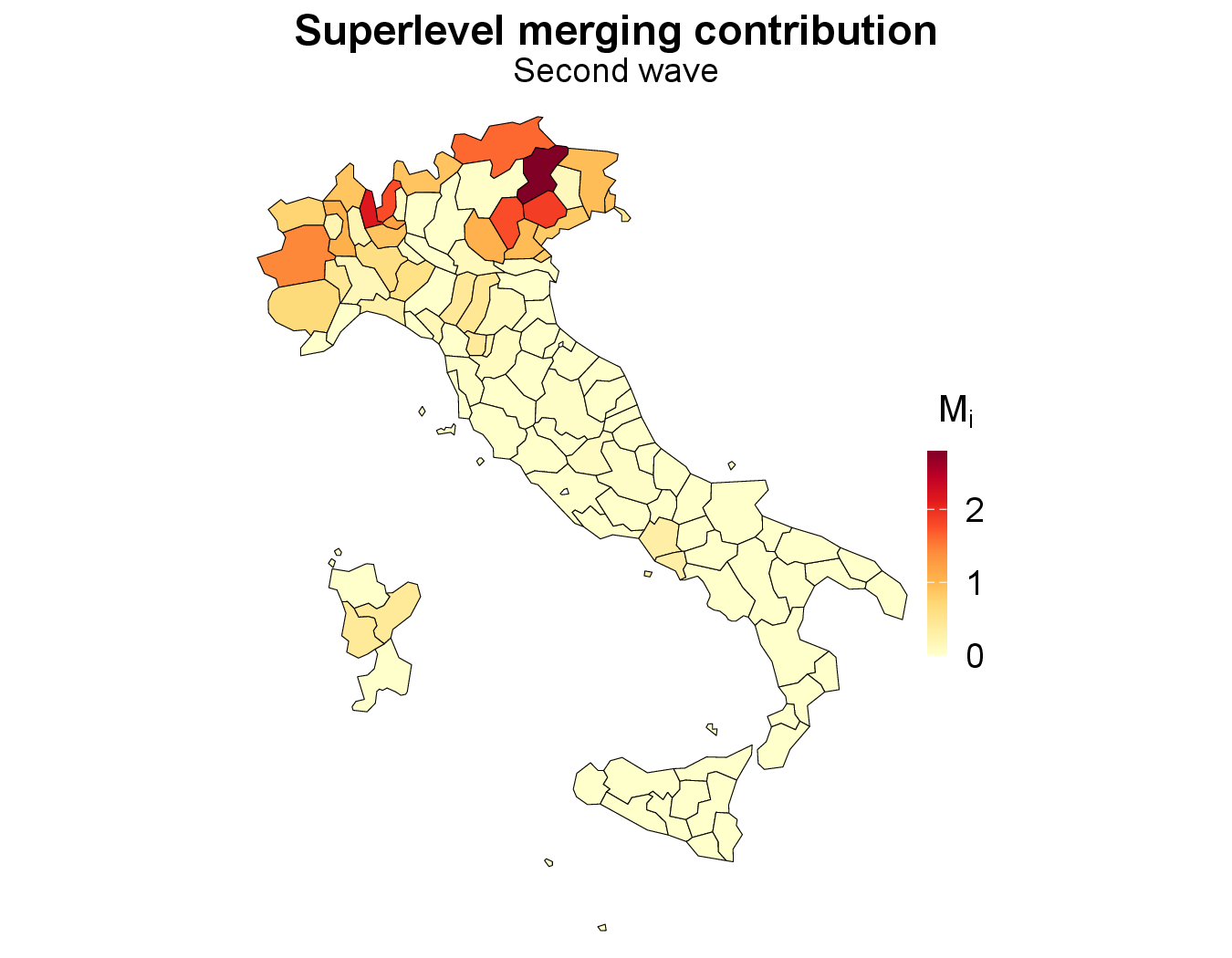}}
    \subfloat[]{\includegraphics[width=8cm,angle=0]{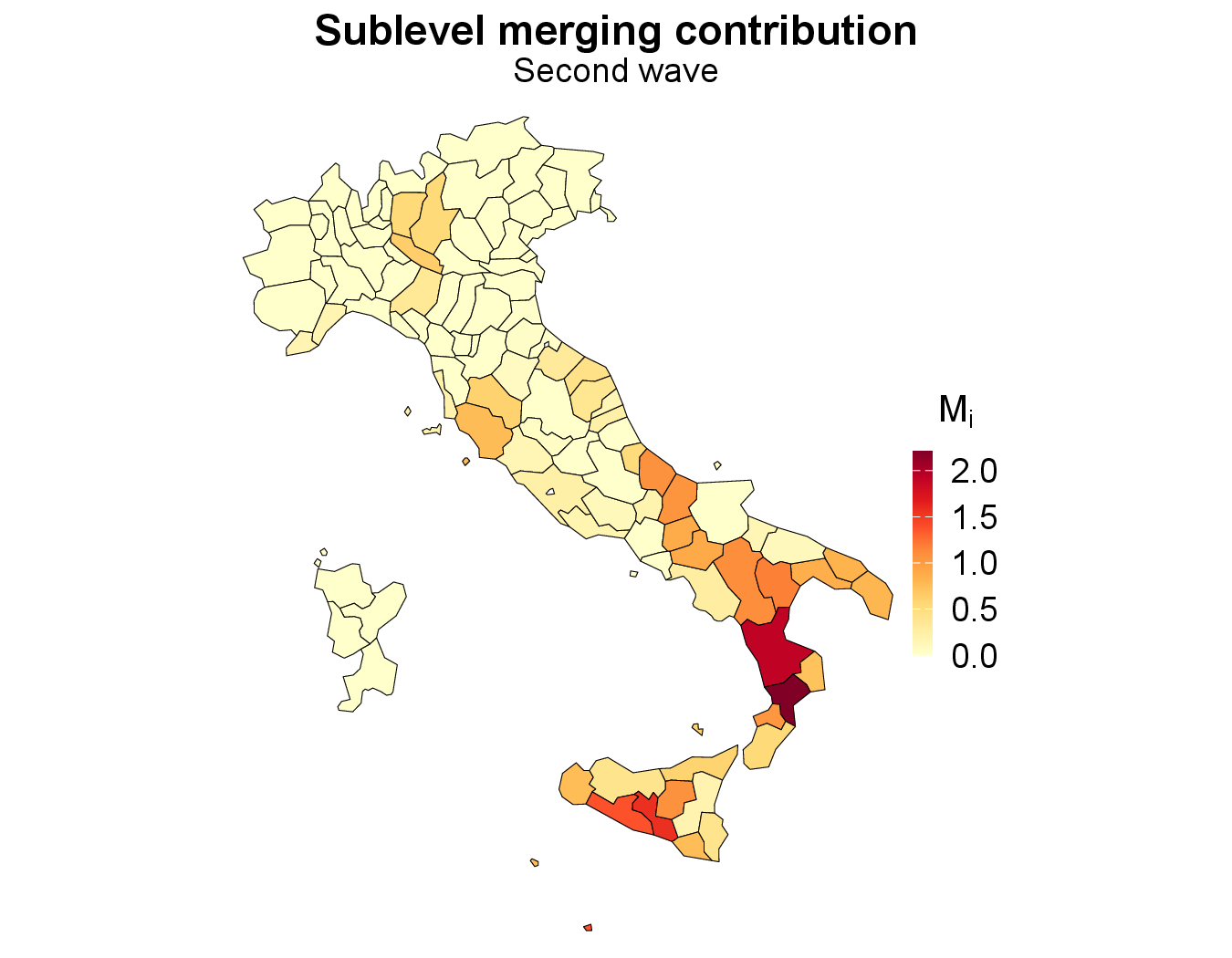}}
    \caption{Integrated local merging contributions during the two epidemic waves. Panels (a) and (c) show $M_i(\mathbf{z})$ for the first and second waves, respectively. Panels (b) and (d) show $M_i(-\mathbf{z})$. Larger values indicate greater cumulative participation in non-redundant component fusions.}
    \label{fig:local-merging-contributions}
\end{figure}

The global analysis shows significant departures from spatial randomness for both high- and low-incidence areas, but it does not identify which provinces contribute most to these connectivity patterns. We therefore turn to the local merging indicators and their conditional tests. Figure~\ref{fig:local-merging-contributions} shows the integrated merging contributions. Larger values of $M_i(\mathbf{z})$ indicate stronger cumulative participation in the fusions connecting high-incidence superlevel regions, whereas larger values of $M_i(-\mathbf{z})$ indicate the analogous role for low-incidence sublevel regions. The largest superlevel contributions are concentrated mainly in Northern Italy, while larger sublevel contributions are more common in southern and insular provinces. These maps provide the continuous local description of connectivity before conditional inference is considered.

\begin{figure}[htbp]
    \centering
    \subfloat[]{\includegraphics[width=8cm,angle=0]{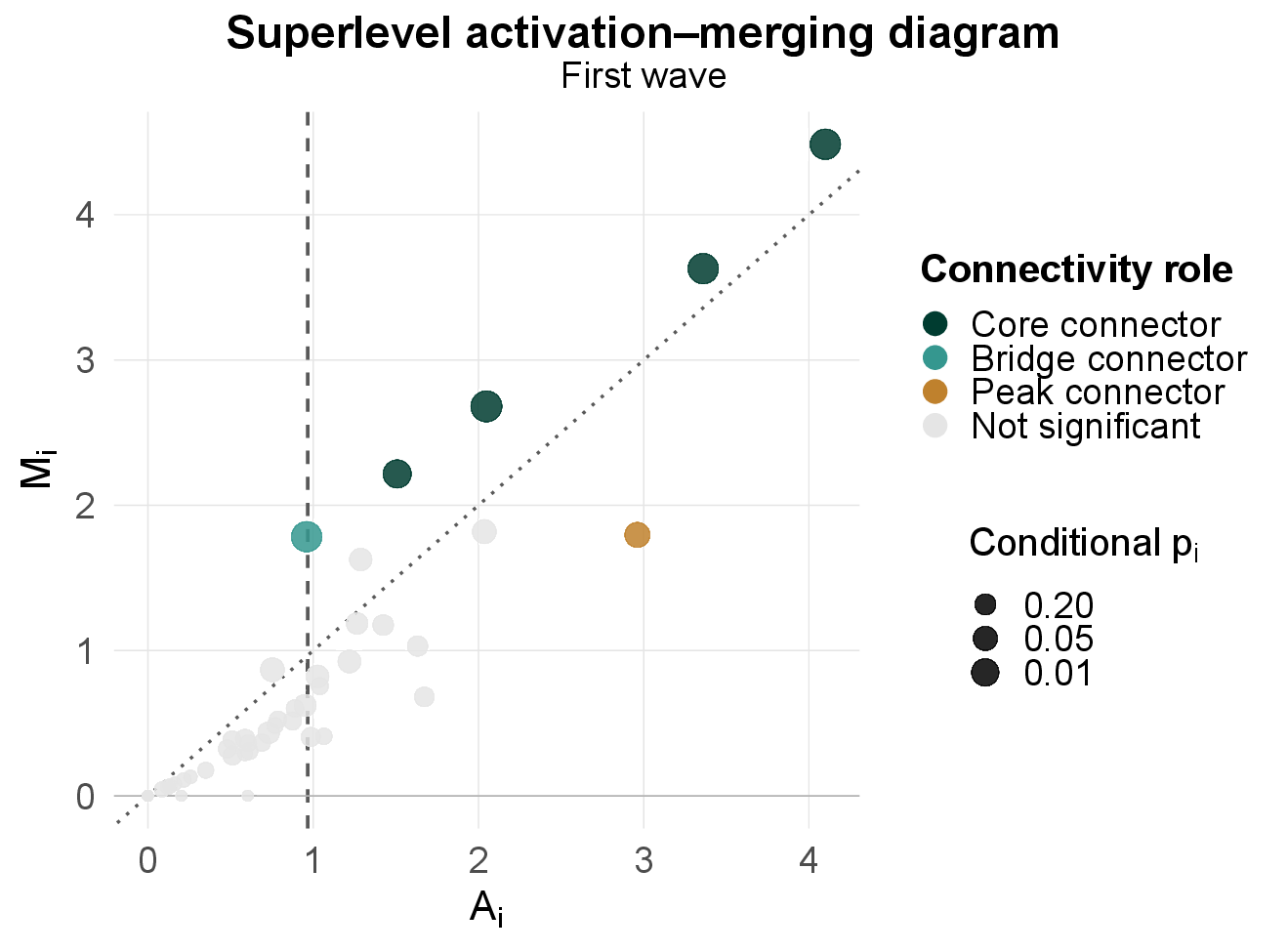}}
    \subfloat[]{\includegraphics[width=8cm,angle=0]{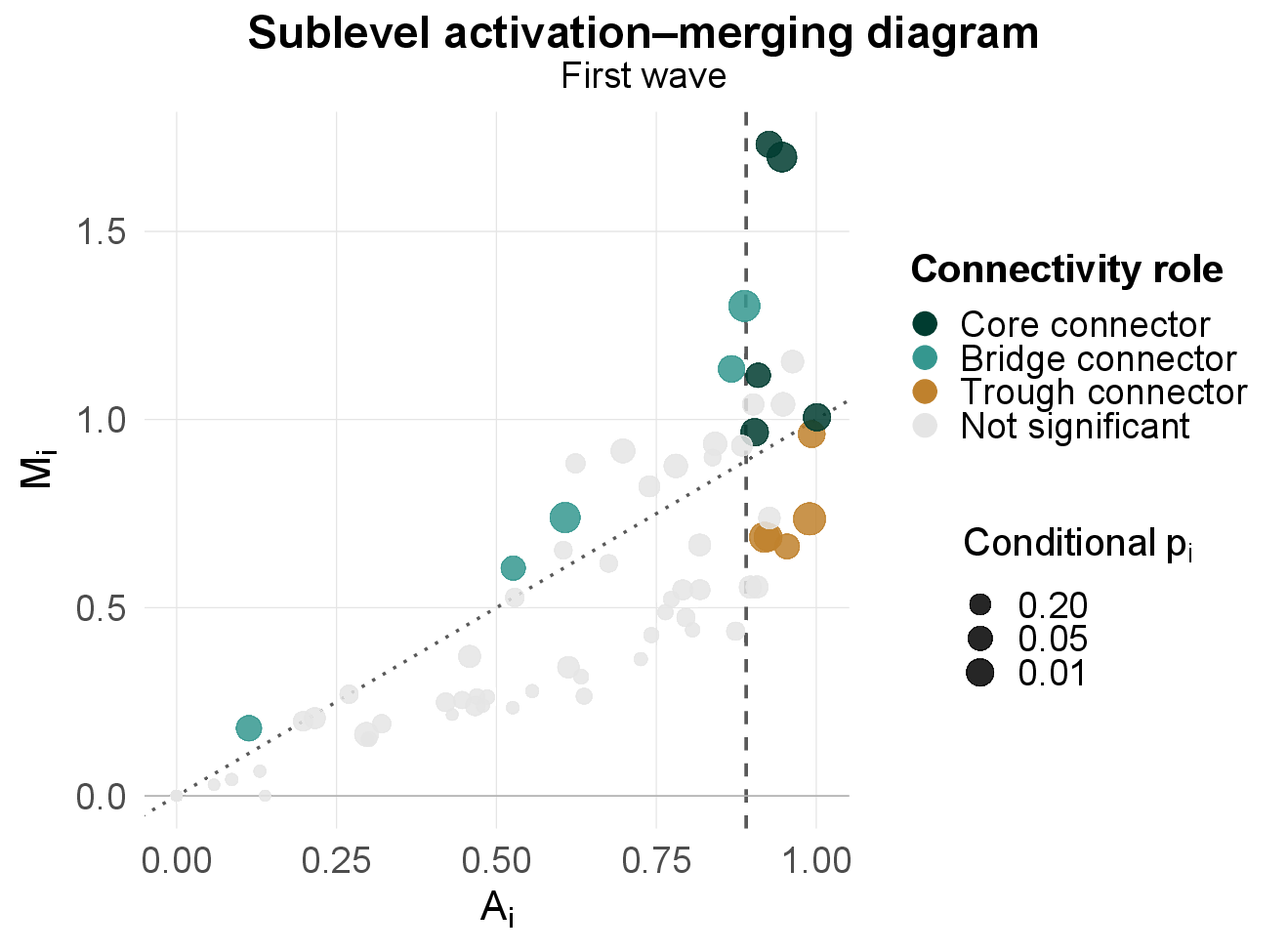}}\\
    \subfloat[]{\includegraphics[width=8cm,angle=0]{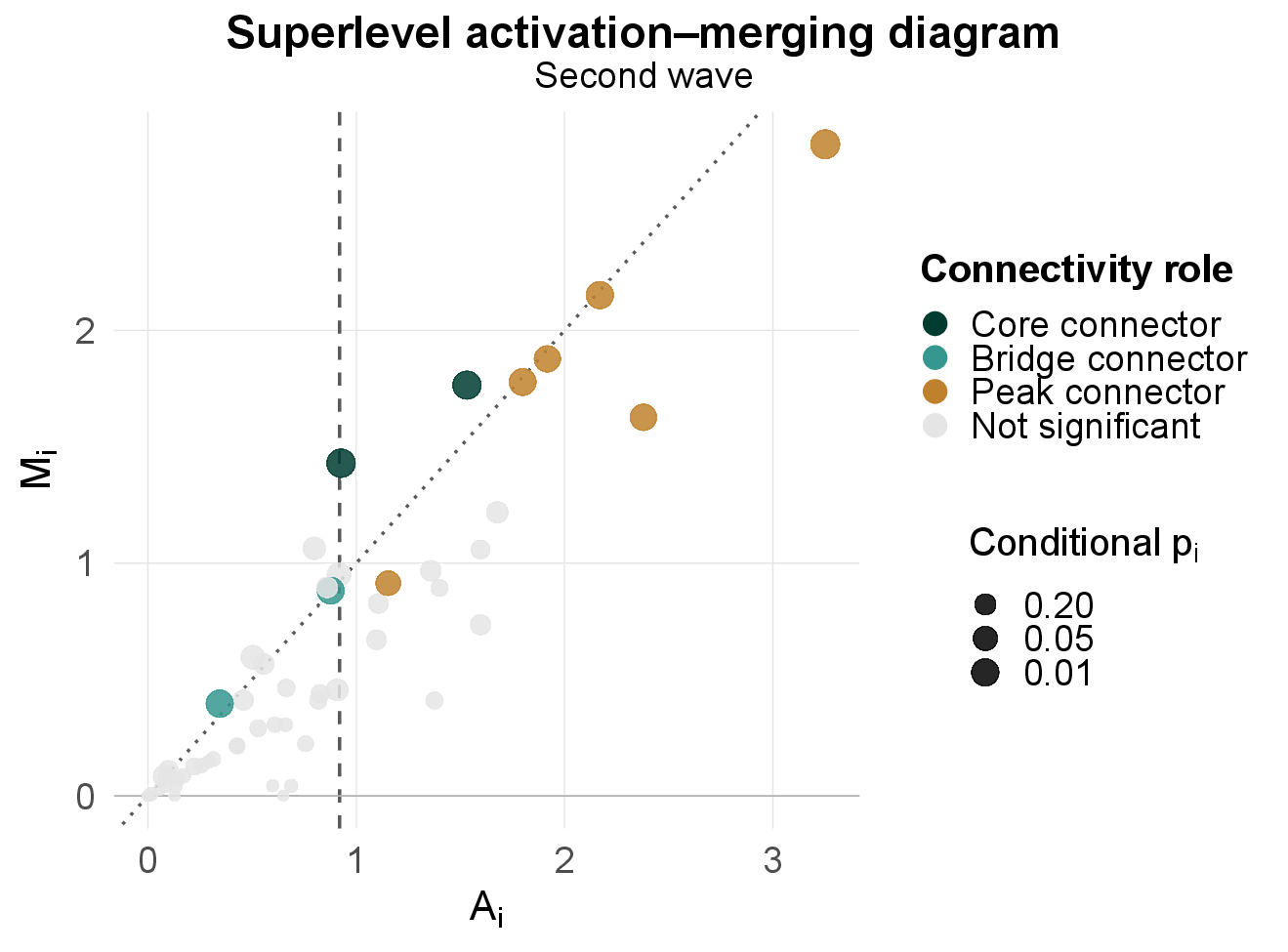}}
    \subfloat[]{\includegraphics[width=8cm,angle=0]{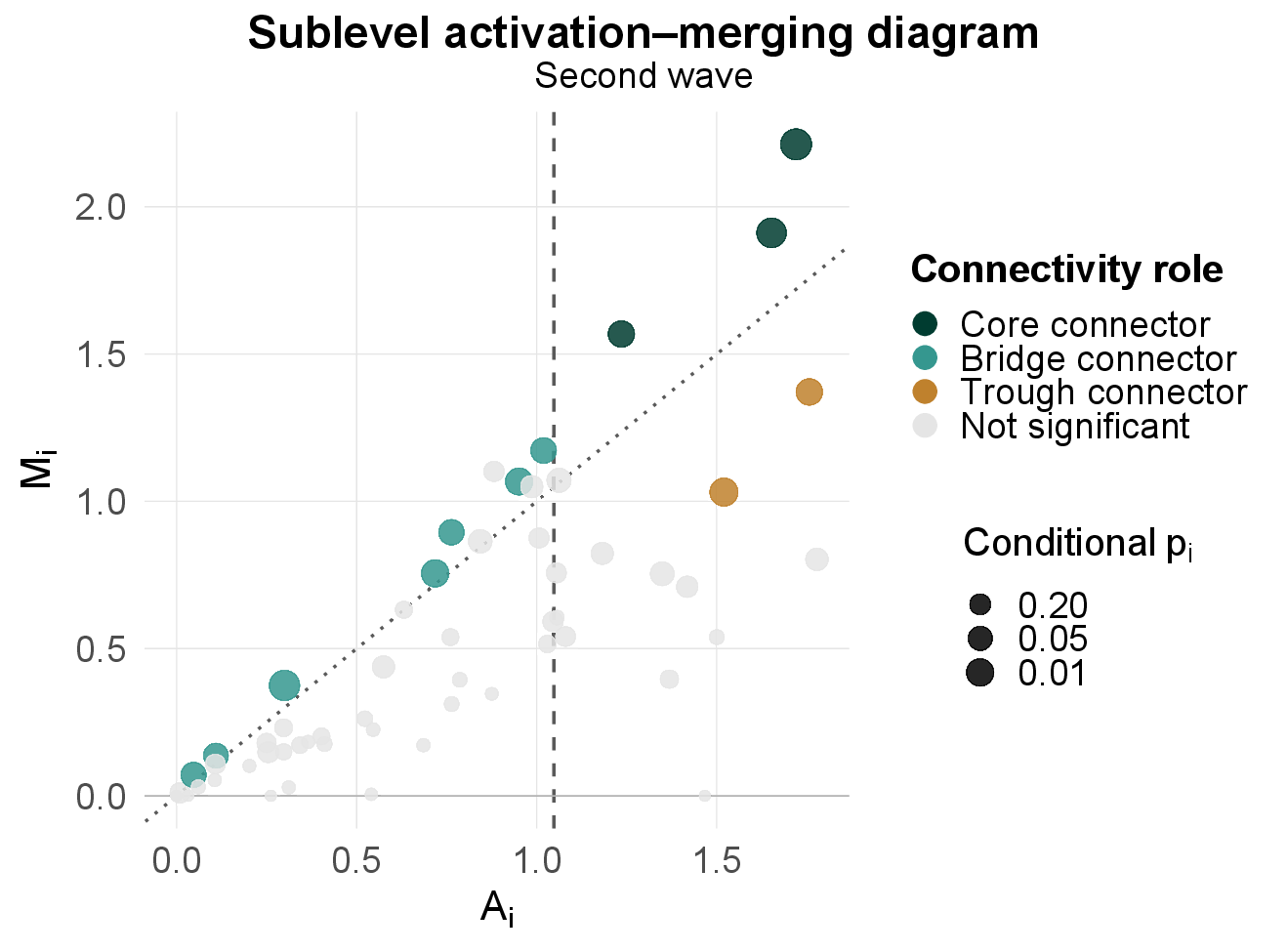}}
    \caption{Activation--merging diagrams for the two epidemic waves. Panels (a) and (c) show $A_i(\mathbf{z})$ against $M_i(\mathbf{z})$, and panels (b) and (d) show $A_i(-\mathbf{z})$ against $M_i(-\mathbf{z})$. Points are colored by the local connectivity profile and sized according to $-\log_{10}(p_i)$. The dotted diagonal represents $M_i=A_i$, and the dashed vertical line marks the empirical tail cut-off with $\eta=0.15$.}
    \label{fig:activation-merging-scatter}
\end{figure}

Figure~\ref{fig:activation-merging-scatter} displays the integrated activation and merging contributions jointly. Points above the diagonal satisfy $M_i>A_i$ and therefore have negative net contributions, whereas points below the diagonal have positive net contributions. The position relative to the diagonal is descriptive and should not be interpreted as a significance criterion; conditional significance is represented separately by point size, which is proportional to $-\log_{10}(p_i)$. The vertical dashed line marks the empirical activation cut-off corresponding to $\eta=0.15$. The resulting cut-offs are 0.97 and 0.89 for the superlevel and sublevel analyses of the first wave, and 0.92 and 1.05 for the corresponding analyses of the second wave.

\begin{figure}[htbp]
    \centering
    \subfloat[]{\includegraphics[width=8cm,angle=0]{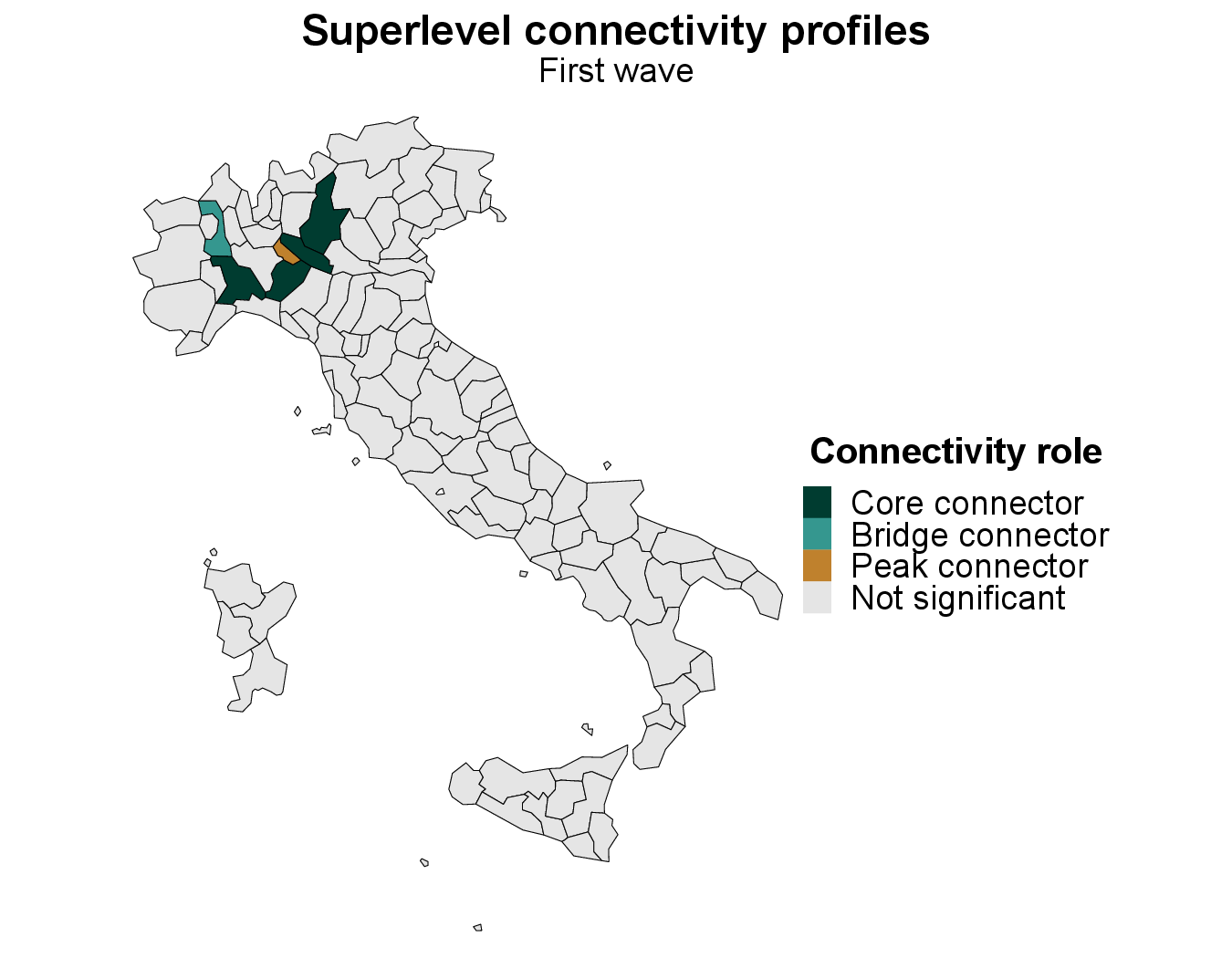}}
    \subfloat[]{\includegraphics[width=8cm,angle=0]{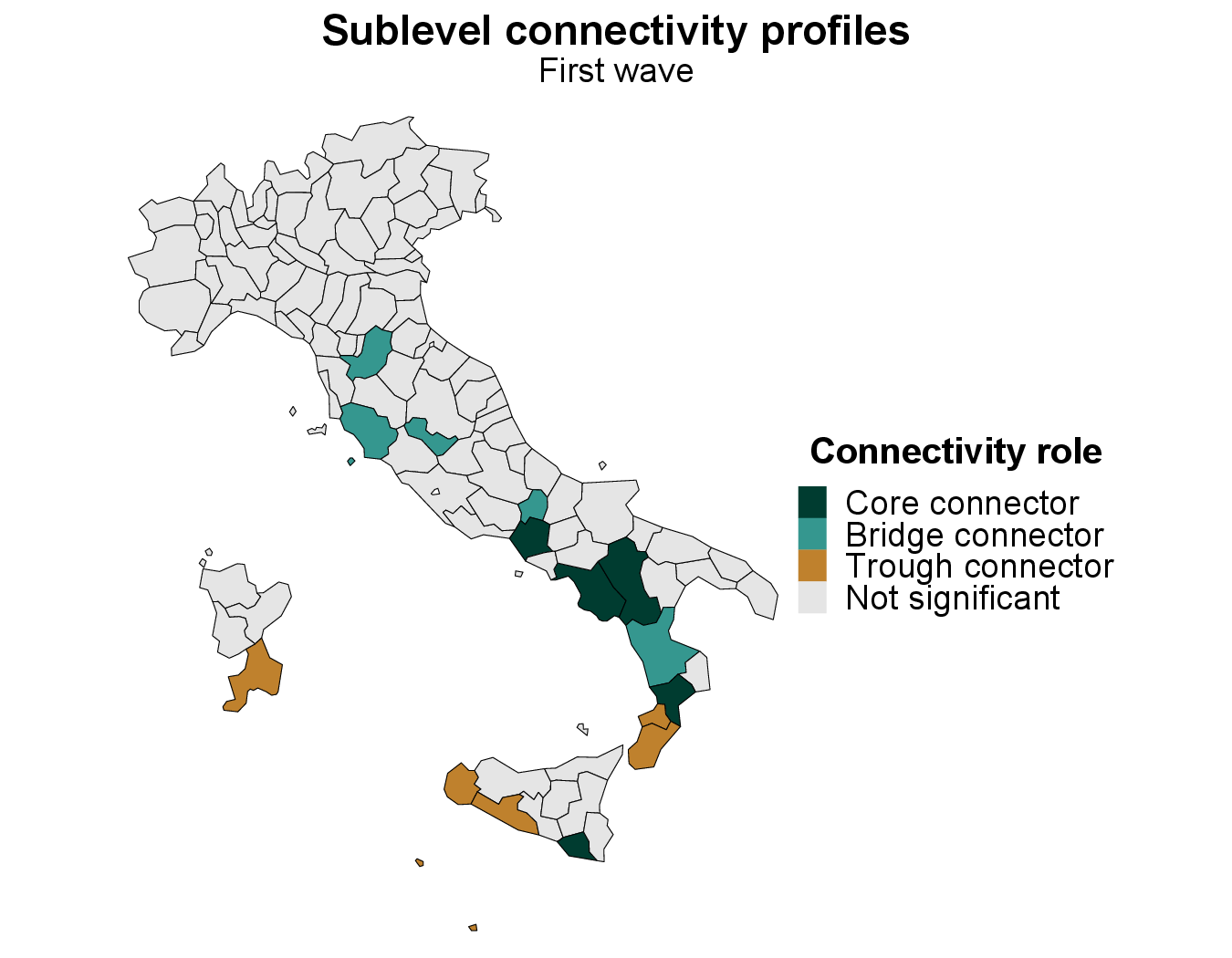}}\\
    \subfloat[]{\includegraphics[width=8cm,angle=0]{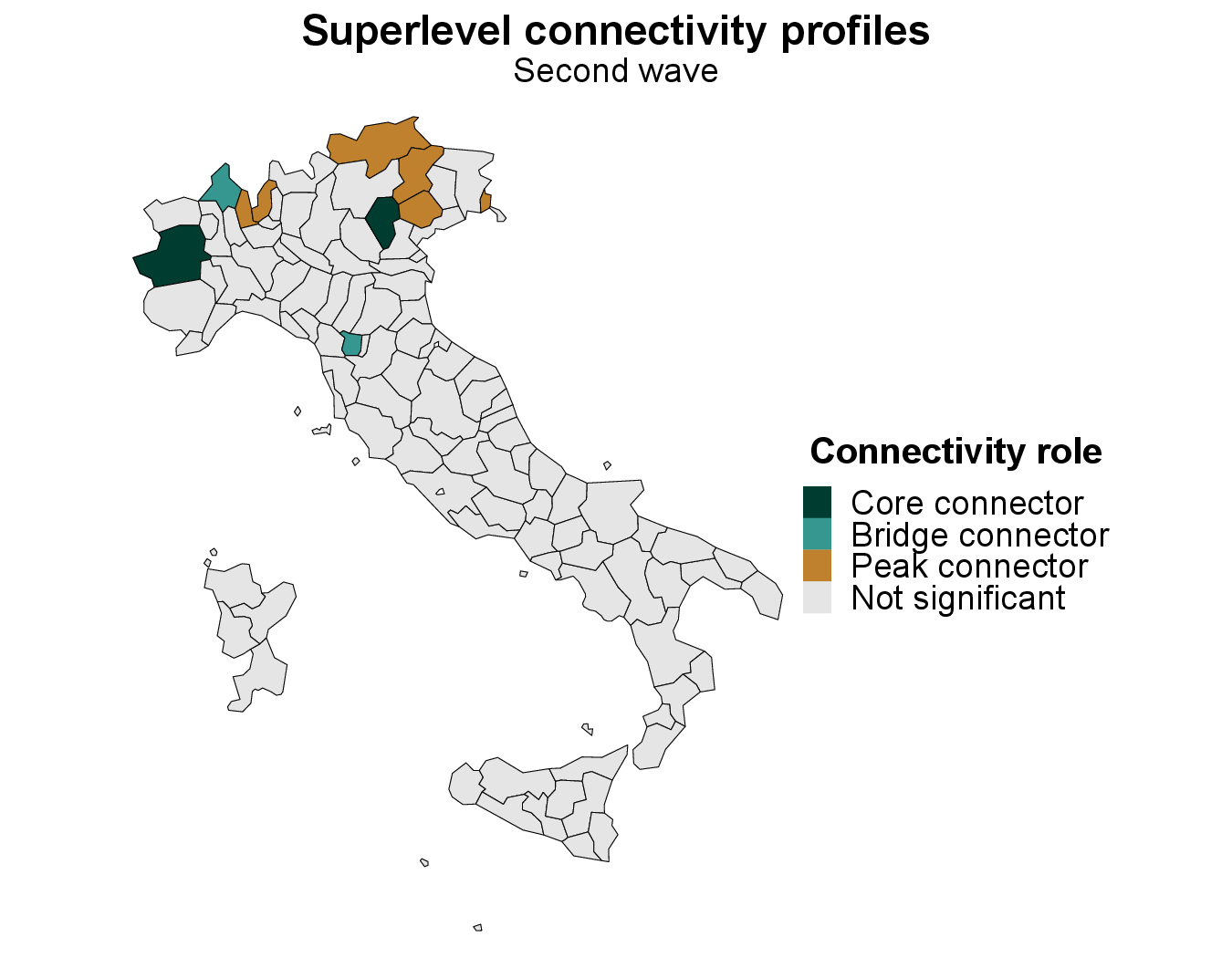}}
    \subfloat[]{\includegraphics[width=8cm,angle=0]{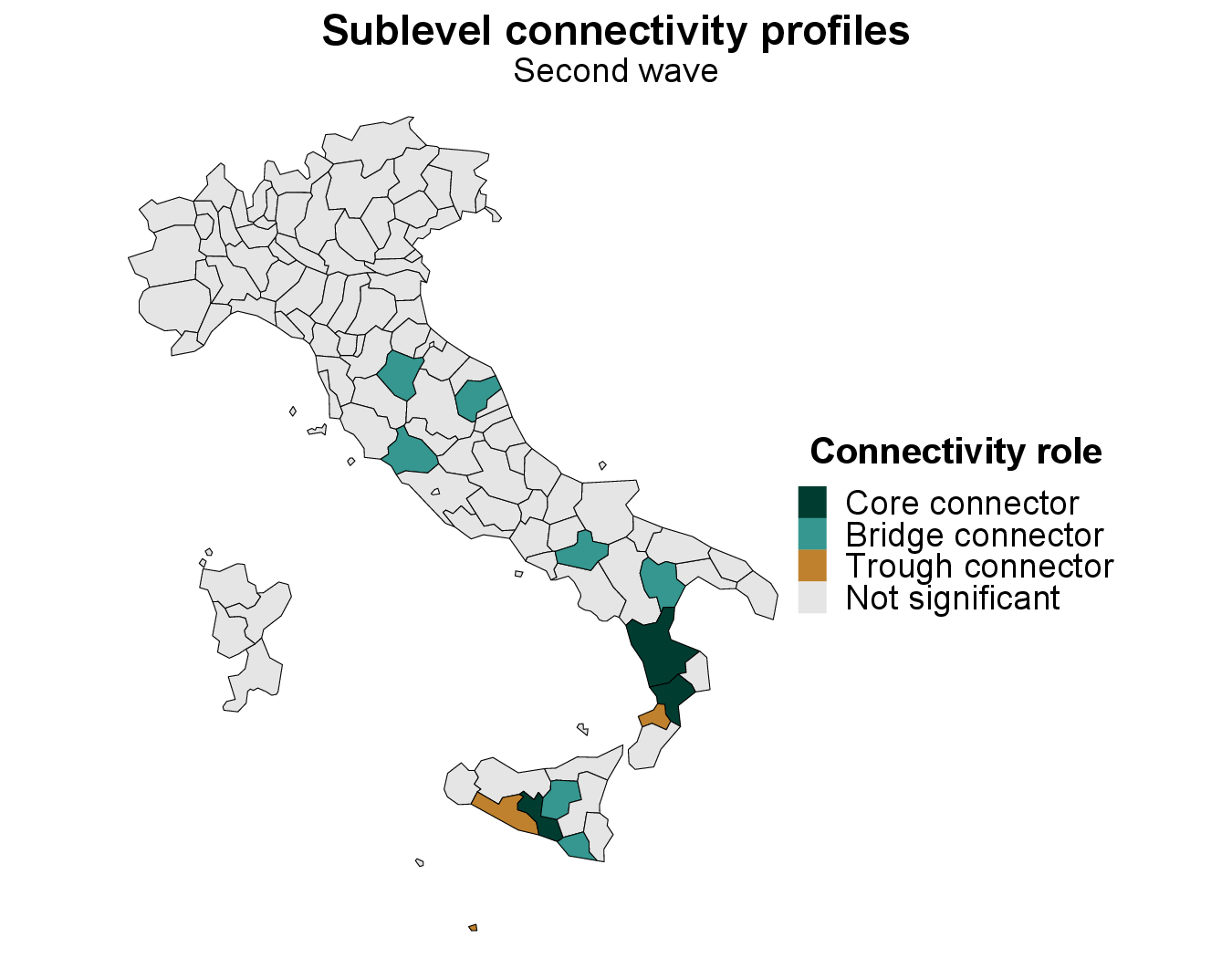}}
    \caption{Local activation--merging connectivity roles for the two epidemic waves. Panels (a) and (c) show the classification for the superlevel analysis, while panels (b) and (d) show the classification corresponding to the sublevel analysis. High activation is defined by the empirical 15\% tail of $\mathbf{z}$ or $-\mathbf{z}$, depending on the direction of analysis.}
    \label{fig:local-connectivity-profiles}
\end{figure}

The connectivity roles in Figure~\ref{fig:local-connectivity-profiles} use unadjusted conditional $p$-values below 0.05. In the first-wave superlevel analysis, six provinces meet this criterion. Brescia, Piacenza, Cremona, and Alessandria are core connectors, Vercelli is a bridge connector, and Lodi is a peak connector. Four of these provinces---Brescia, Piacenza, Cremona, and Vercelli---remain significant after Benjamini--Hochberg adjustment, with adjusted $p$-values of 0.0398. These results identify a compact set of northern provinces with strong merging roles in the highly connected first-wave superlevel structure.

Turning to the second wave, the superlevel analysis contains ten connectors at the unadjusted 0.05 level. Torino and Vicenza are core connectors; Pistoia and Verbano-Cusio-Ossola are bridge connectors; and Belluno, Varese, Como, Bolzano-Bozen, Treviso, and Gorizia are peak connectors. None remains significant after false discovery rate (FDR) adjustment. Compared with the first wave, the superlevel local signal is distributed over a larger and more heterogeneous set of northern and north-eastern provinces, which is consistent with the broader spatial pattern of elevated incidence during the second wave.

The sublevel analyses shift attention from connected high-incidence structures to the organization of provinces below the within-wave reference. In the first wave, 15 provinces meet the unadjusted 0.05 criterion. Caserta, Salerno, Ragusa, Potenza, and Catanzaro are core connectors; Cosenza, Terni, Isernia, Firenze, and Grosseto are bridge connectors; and Trapani, Vibo Valentia, Reggio di Calabria, Agrigento, and Cagliari are trough connectors. The second-wave sublevel analysis identifies 12 connectors at the unadjusted 0.05 level: Catanzaro, Cosenza, and Caltanissetta are core connectors; Macerata, Ragusa, Enna, Matera, Avellino, Arezzo, and Viterbo are bridge connectors; and Vibo Valentia and Agrigento are trough connectors. No sublevel-analysis result remains significant after FDR adjustment in either wave.

\begin{figure}[htbp]
    \centering
    \subfloat[]{\includegraphics[width=8cm,angle=0]{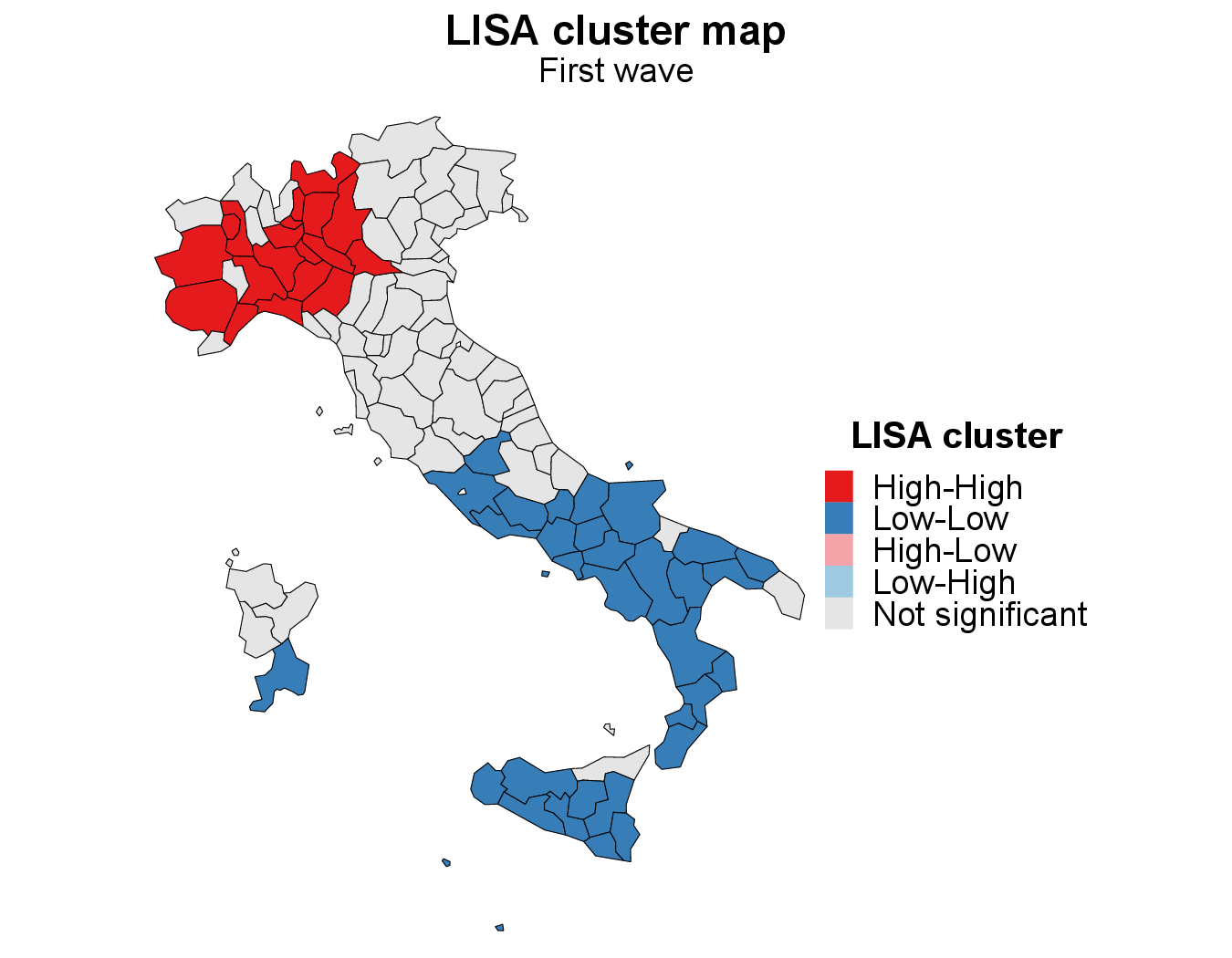}}
    \subfloat[]{\includegraphics[width=8cm,angle=0]{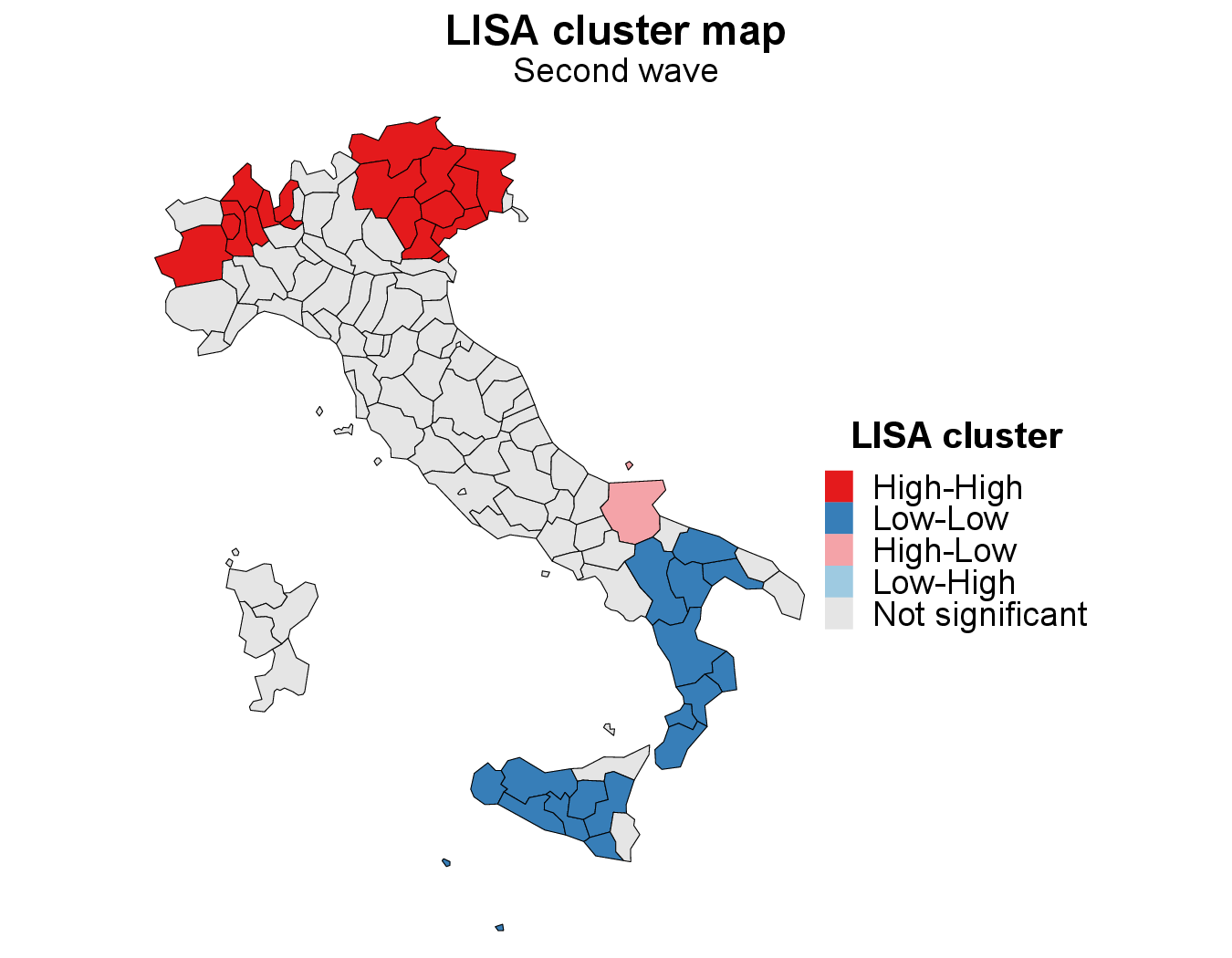}}
    \caption{Local Moran's $I$ cluster maps for the two epidemic waves. Panels (a) and (b) show the first and second waves, respectively. The classifications use unadjusted two-sided permutation $p$-values below 0.05.}
    \label{fig:lisa-clusters-waves}
\end{figure}

Comparison with Local Moran's $I$ (Figure \ref{fig:lisa-clusters-waves}) shows both overlap and differences between local association and local connectivity. All six first-wave superlevel connectors are also classified as High--High areas at the unadjusted 0.05 level. In the first-wave sublevel analysis, however, Terni, Firenze, and Grosseto have significant connectivity roles without being Low--Low areas at the unadjusted 0.05 level. In the second-wave superlevel analysis, Pistoia and Gorizia are connectors but are not High--High at the unadjusted 0.05 level, while in the second-wave sublevel analysis the same occurs for Macerata, Avellino, Arezzo, and Viterbo relative to the Low--Low category. These cases illustrate that an area can contribute to the connectivity of threshold-defined regions without showing significant local similarity under Local Moran's $I$.

The local results therefore refine the global connectivity patterns rather than reproducing standard local association maps. The continuous $M_i$ values remain the main local outputs, while the connector profiles provide a compact summary of unadjusted conditional evidence and activation level. The four false-discovery-rate significant first-wave superlevel results give stronger inferential support for part of the northern connectivity pattern, whereas the remaining connectivity roles found should be interpreted with more caution.

\section{Discussion and conclusions}
\label{sec:discussion}

This paper introduces global and local indicators of spatial connectivity for areal data. The aim is to complement conventional measures of spatial association with summaries that describe whether values beyond a reference level form connected regions over an areal graph. Moran's $I$, Geary's $C$, and local indicators of spatial association describe relationships between neighboring values, while the proposed framework focuses on the number and organization of connected components in threshold-induced spatial graphs.

At the global level, the Betti-0 curve records the number of connected components as the threshold moves away from the reference level. Superlevel and sublevel structures are handled with the same construction by analyzing $\mathbf{z}$ and $-\mathbf{z}$, respectively. A random-relabeling test compares the observed curve with spatially exchangeable configurations. The complete curve indicates where departures occur across thresholds, while the integrated discrepancy provides a scalar global test statistic.

At the local level, the method is built on an allocation of component fusions. When an area enters the filtration, each already active component that it touches generates one merging unit. Half of that unit is assigned to the entering area, and the other half is divided equally among its neighbors in the component being joined. The normalization is applied separately to each component, so an area touching $q_i$ active components generates $q_i$ merging units. This yields a well-defined local allocation when active vertex values are distinct and preserves the exact identity between the integrated Betti-0 curve and the sum of the local activation and merging contributions.

The application to Italian COVID-19 incidence identifies significant global superlevel and sublevel connectivity during both epidemic waves. The largest departure corresponds to high incidence in the first wave, consistent with the concentration of the early outbreak in Northern Italy. The local indicators identify six superlevel connectors at the unadjusted 0.05 level in this period, four of which remain significant after FDR adjustment. The other analyses contain broader sets of connectors at the unadjusted level but no FDR-adjusted local signals. Comparison with Local Moran's $I$ also shows that local association and local connectivity can select different areas, particularly in the sublevel analyses and in the second-wave superlevel analysis.

Although the application concerns infectious-disease incidence, the methodology is not tied to epidemiological outcomes. In environmental studies, the same approach can be used to examine whether pollutant exceedances, high temperatures, drought conditions, wildfire risk, or water-quality deficits form coherent spatial structures or separated hotspots. Ecological applications may focus on connected areas of habitat degradation, biodiversity loss, ecosystem stress, or other threshold-defined indicators. In these settings, the local merging contribution can identify areas that are important for maintaining the connectivity of the observed pattern, providing information that is different from pairwise spatial association.

Several limitations should be kept in mind. First, connectivity depends on the adjacency graph, and alternative scientifically plausible graphs may lead to different results. The choice of reference level \(b\) and scale \(s\) also affects the interpretation and units of the indicators and should therefore be justified for each application. In addition, the proposed inferential procedures rely on spatial exchangeability, and alternative resampling schemes may be needed when this assumption is not appropriate. Finally, the current local decomposition assumes that the positive entries of \(\mathbf z\) are distinct, so that areas enter the superlevel filtration one at a time. This condition is expected to hold for many continuously valued outcomes, whereas exact ties, for example due to substantial rounding, would require a simultaneous treatment of tied areas.

The present work focuses on Betti-0 because connected components have a direct interpretation for areal maps and permit an exact local decomposition. Further research could consider other topological summaries, including Betti-1 for one-dimensional holes, as well as extensions to spatio-temporal settings and goodness-of-fit analysis. Overall, the proposed framework shows how a simple topological quantity can be translated into global and local tools for studying spatial connectivity in environmental, ecological, epidemiological, and other areal data.

\appendix

\section{Proofs}
\label{app:proofs}

\subsection*{Proof of Proposition~\ref{prop:activation-merging-decomposition}}
Order the active vertices with $z_j\geq\lambda$ from largest to smallest value. Immediately before area $j$ is added, the already active graph is $G_j^+$. By definition, $j$ has neighbors in exactly $q_j$ connected components of this graph. If $j$ is first regarded as a singleton, adding its edges to already active neighbors replaces these $q_j+1$ components by one component, so the corresponding change in the component count is $1-q_j$. Summing over all vertices active at threshold $\lambda$ gives
\[
\beta_0(\lambda;\mathbf{z})
=
\sum_{j:z_j\geq\lambda}(1-q_j)
=
|V_\lambda(\mathbf{z})|-\sum_{j:z_j\geq\lambda}q_j.
\]
The activation identity follows directly from the definition of $a_i(\lambda)$. For each fusion $(j,C)$, the allocation weights satisfy
\[
\sum_{i\in V}\omega_{j\to i}(C)
=
\frac12+k_j(C)\frac{1}{2k_j(C)}=1.
\]
Therefore, using the definition of $m_i(\lambda)$ and interchanging the finite sums,
\[
\begin{aligned}
\sum_{i\in V}m_i(\lambda)
&=
\sum_{i\in V}
\sum_{j:z_j\geq\lambda}
\sum_{C\in\mathcal C_j}
\omega_{j\to i}(C)\\
&=
\sum_{j:z_j\geq\lambda}
\sum_{C\in\mathcal C_j}
\sum_{i\in V}\omega_{j\to i}(C)\\
&=
\sum_{j:z_j\geq\lambda}
\sum_{C\in\mathcal C_j}1\\
&=
\sum_{j:z_j\geq\lambda}q_j.
\end{aligned}
\]
Substituting these two identities into the preceding expression for $\beta_0(\lambda)$ gives the activation--merging representation, and the equality involving $\ell_i(\lambda)$ follows from its definition.

\subsection*{Proof of Proposition~\ref{prop:random-relabel}}
By Proposition~\ref{prop:activation-merging-decomposition}, for each $k$,
\[
\beta_0\bigl(\lambda;\mathbf{z}^{(k)}\bigr)
=
\sum_{i\in V}a_i\bigl(\lambda;\mathbf{z}^{(k)}\bigr)
-
\sum_{i\in V}m_i\bigl(\lambda;\mathbf{z}^{(k)}\bigr).
\]
Every relabeling contains the same multiset of values, so the total number of active vertices is the same for all $k$ at a fixed threshold. Subtracting the common mean curve therefore cancels the activation terms, leaving exactly the negative difference between the corresponding total merging contribution and its mean over the $R+1$ configurations.

\end{document}